\documentclass[twocolumn]{aastex701}
\usepackage[dvipsnames,svgnames]{xcolor}
\usepackage{enumitem}

\begin{document}
\title{Beyond the blur: Radial binning trumps stellar migration in resolved star formation histories}


\author[0000-0002-8085-7578]{Charlotte Olsen$^*$}
\affiliation{NYC College of Technology,
City University of New York,
NY, USA}
\affiliation{LSST Discovery Alliance,
Catalyst fellowship}
\email[show]{charlotte.olsen47@citytech.cuny.edu}

\author[0000-0001-5576-0144]{Charlotte Welker$^*$}
\affiliation{NYC College of Technology,
City University of New York,
NY, USA}
\affiliation{Graduate Center,
City University of New York,
NY USA}
\email[]{cwelker@citytech.cuny.edu}

\author[]{Ena Chia$^*$}
\affiliation{NYC College of Technology,
City University of New York,
NY, USA}
\email[]{ena.a.chia@gmail.com}

\author[0000-0002-1685-5818]{Anna C Wright}
\affiliation{Center for Computational Astrophysics,
Flatiron Institute, New York
NY, USA}
\email[]{awright@flatironinstitute.org}

\correspondingauthor{Charlotte Olsen}

\begin{abstract}
Spatially resolved galaxy properties are routinely derived from integral-field spectroscopy and increasingly from multi-band photometry as a wealth of photometric data become available in large surveys like LSST/Rubin. Spatially resolved star formation histories (SFHs) offer a detailed view of galaxy growth and quenching by tracing the timing and location of star formation within a galaxy up to one billion years in the past. However, stellar migration between spaxels may strongly bias these reconstructions. Quantifying this bias remains challenging as few Milky-Way mass simulations fully resolve initial stellar dispersions or migration rates. We use the well-validated h277 Milky-Way-mass cosmological zoom simulation, which resolves both via explicitly tying star formation to the non-equilibrium abundance of $H_2$, to assess the quality of specific star formation rates (sSFR) radial profiles inferred from resolved SFHs against actual values from stellar particle tracking. Across several evolutionary windows spanning a range of stellar activity stages from quiescent to star-bursting, we find that the adaptive radial binning imposed by S/N constraints in photometric SED fitting actually mitigates most of the blurring/mixing effects of stellar migration on inferred radial profiles up to lookback times of 1.5 Gyr. This allows for proper recovery of not only the average sSFR but also its radial slope, its relative changes in scatter across radial annuli and even approximate localization of low-z star-forming clumps. Only galaxies having undergone a major merger across the considered lookback time display discrepant, unreliable sSFR profiles, hence should not be considered for such reconstructions. 
\end{abstract}

\section{Introduction}

Modern observational studies of galaxy evolution typically rely on the analysis of statistical samples of galaxies. However, these are constrained to adopt either a high-sampling statistical approach in which global properties and scaling relations are inferred from large, often photometric surveys \citep[e.g.,][]{York2000, Brinchmann2004, Grogin2011, Koekemoer2011}, at the expense of spectral and/or spatial resolution, or a spatially and spectrally resolved approach in which internal physical mechanisms governing growth and quenching are constrained through high-resolution observations but at the expense of sampling \citep[integral-field spectroscopy; e.g.,][]{Sanchez2012, Croom12, Bundy2015}. In the resolved regime, prior work has characterized the buildup and suppression of star formation by measuring the star formation rate (SFR) or specific star formation rate (sSFR) as a function of galactocentric radius via radial profiles \citep[e.g.,][]{Tacchella2015, Nelson2016, Belfiore2018, Ellison2018}. Under this interpretation, elevated SFR traces ongoing stellar mass assembly, whereas declining or centrally/externally suppressed star formation is associated with progression toward quiescence. Accordingly, ``inside-out'' versus ``outside-in'' growth or quenching is commonly characterized through the sign and magnitude of the radial SFR/sSFR gradient \citep[e.g.,][]{Perez2013, GonzalezDelgado2014, Tacchella2015, Belfiore2018}.

Additional constraints can be derived by inferring stellar ages and formation times, which enables the reconstruction of star formation histories (SFHs). For sub-regions within nearby galaxies, SFHs are most robustly inferred by modeling color--magnitude diagrams (CMDs) in fields where individual stars can be resolved \citep[e.g.,][]{Dolphin2002, Tolstoy2009}. The maximum distance at which CMD-based SFHs can be recovered is set by observational depth and crowding limitations, restricting such analyses largely to the Local Volume \citep[e.g.,][]{dalcanton}, although JWST is extending CMD-based SFHs to $\sim$10--20\,Mpc \citep[e.g.,][]{Bortolini2024, Lee2026}. For more distant systems, SFHs are instead derived from integrated-light spectroscopy or integral-field unit (IFU) observations \citep[e.g.,][]{CidFernandes2005, walcher, Perez2013, GonzalezDelgado2014}. More recently, substantial effort has focused on exploiting the extensive multiwavelength imaging available across large galaxy samples and cosmic time to infer spatially resolved SFHs by fitting spectral energy distributions (SEDs) for subgalactic regions \citep{Wuyts2012, Hemmati2020, GimenezArteaga2023}. These regions may be defined on a per-pixel basis or via algorithmic segmentation of adjacent pixels with similar photometric properties \citep[e.g.,][]{Cappellari2003, Wuyts2012, Abdurrouf2022a, Abdurrouf2022, Abdurrouf2025}. At the highest redshifts, however, ``resolved'' SFH constraints may be limited in practice to coarse structural components (e.g., bulge--disk decompositions) rather than genuinely fine spatial substructure \citep[e.g.,][]{Lang2014}.

Whether the analysis is based on resolved stellar populations or on galaxy-integrated photometry, the reconstruction of a star-formation history (SFH) generally relies on the assumption that the relevant stellar populations formed in situ. For a galaxy-wide SFH, this approximation is often adequate because the photometric and spectroscopic properties of old stellar populations are broadly similar whether those stars formed locally or were accreted through mergers \citep[e.g.,][]{conroyrev}. In contrast, spatially resolved SFHs introduce an additional source of systematic uncertainty by implicitly assuming that all stars presently observed within a given spatial footprint were also born there. This assumption of negligible stellar redistribution is not physically realistic: dynamical processes that drive stellar motions and radial migration can mix populations across the disk or halo \citep[e.g.,][]{Roskar2008, SanchezBlazquez2009}. Consequently, stellar migration tends to wash out local SFH signatures, such that the oldest inferred epochs in a resolved SFH may primarily reflect a temporally averaged (or median) SFH of the galaxy rather than the true early-time SFH of the specific region under study. Radial migration has become a catch-all term that covers a number of physical processes that change either the radius of a star's orbit or its eccentricity. \citet{SellwoodBinney2002} showed that transient spiral arms can change a star's angular momentum at corotation without heating its orbit, and \citet{SchonrichBinney2009} formalized the distinction between the two main mechanisms that fall under the definition of radial migration. These are ``churning'', in which the guiding radius of a star changes through corotation resonances with transient spiral arms \citep{SellwoodBinney2002, Roskar2008} or with the bar and overlapping bar--spiral resonances \citep{MinchevFamaey2010, Minchev2011, Minchev2012, Halle2015}, and ``blurring'', in which the guiding radius is unchanged but the orbital eccentricity of a star is increased by non-axisymmetric perturbations such as mergers or satellite interactions \citep{Grand2016, Bird2021}, or by interactions with the arms or bar at their Lindblad resonances \citep{SellwoodBinney2002, Minchev2012}. Accounting for the effects of stellar mixing in the interpretation of resolved SFHs is a longstanding challenge. It is well established that galactic substructure and disk dynamical processes can modify stellar angular momentum, leading to dynamical ``heating'' or ``cooling'' and consequent migration onto different orbits. \citet{Minchev2026} recently investigated the implications of these processes for resolved SFHs, demonstrating that SFHs inferred at a given galactocentric radius may systematically under- or overestimate the underlying stellar populations owing to radial redistribution \citep[see also][]{Bernaldez2026}.

This raises a fundamental question regarding the extent to which radial information is retained in the presence of stellar radial migration, and consequently how reliably radial profiles reconstructed from SFHs can be interpreted. \citet{Olsen2026} reconstructed radial profiles from spatially resolved SFHs for a sample of UVCANDELS \citep{Wang2024UVCANDELS} galaxies and reported an overall decline in the sSFR with radius. However, their analysis was restricted to a lookback time of 1 Gyr due to concerns that radial migration could bias the inferred SFHs at earlier epochs. Despite this limitation, the recovered radial trends were consistent with a subset of systems showing the earliest stages of quenching. Extending this approach to large statistical samples requires quantitative validation of the reconstructed profiles when radial migration is explicitly accounted for. If such profiles are shown to be robust, these methods could be applied to the millions of galaxies targeted by current and forthcoming surveys \citep[e.g.,][]{Spergel2015, Ivezic2019, Euclid2025}, enabling constraints on the timing and spatial progression of star-formation cessation across a range of physical scenarios (e.g. environment and redshift).

This paper is organized as follows: In Section~\ref{sec:methodology} we introduce our methodology including a description of the simulation we use and two binning schemes motivated by observations. In Section~\ref{sec:radial_profiles} we demonstrate our ability to reconstruct radial profiles from resolved SFHs, and in Section~\ref{sec:migration} we analyze the migration from bin to bin before discussing our results and providing recommendations for observers in Section~\ref{sec:discussion}. Finally, in Section~\ref{sec:conclusions} we summarize our conclusions.

\section{Methodology}
\label{sec:methodology}


\subsubsection{Overview}

We use the full-physics and robustly tested cosmological hydrodynamical {\bf h277} zoom simulation from the g14 suite \citep{Christensen12,Christensen14,Brooks2025}, which centers on a Milky-Way analog simulated over 13.7 Gyrs with the widely used SPH code GASOLINE \citep{Wadsley04}. It is extensively described in \citet{Christensen12, Christensen14} and \citet{Christensen16} and is publicly available \citep{Brooks2025}. We only provide a brief summary below.

\begin{figure}
    \centering
    \includegraphics[width=0.95\linewidth]{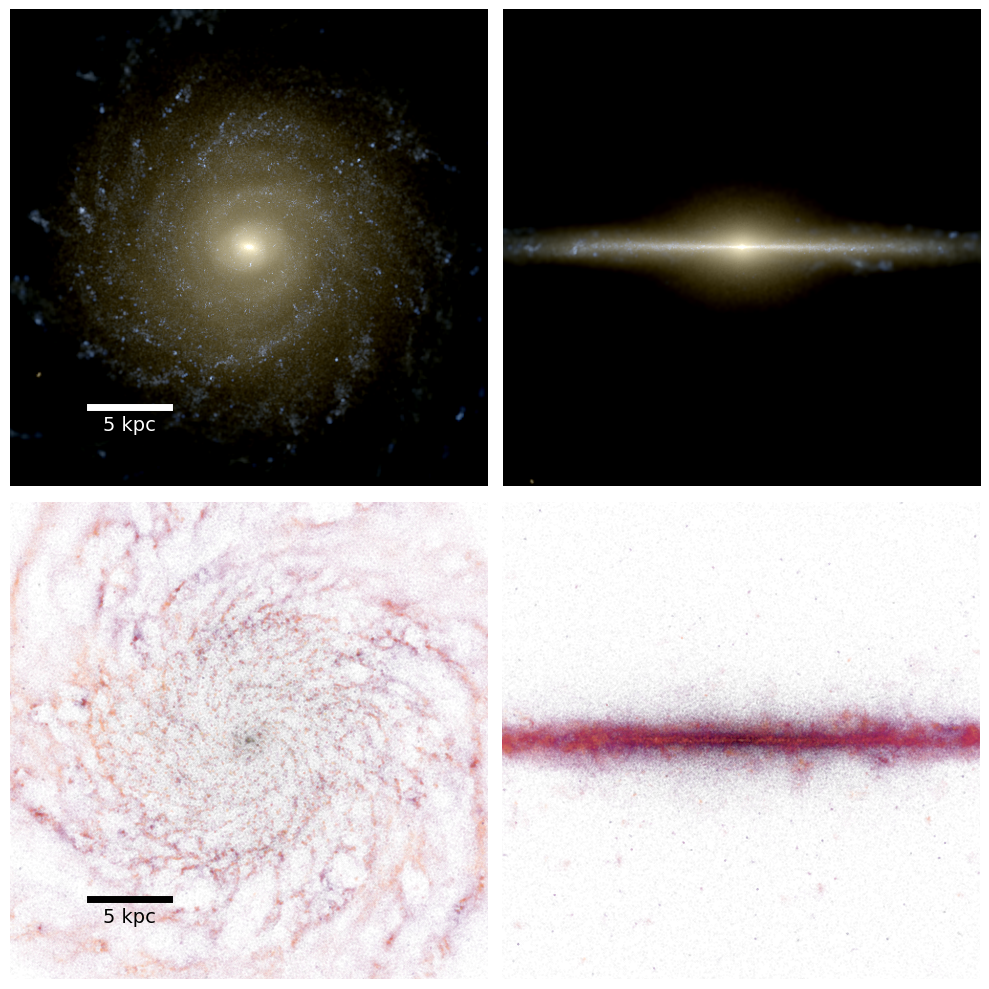}
    \caption{The simulated Milky Way analog h277 at $z=0$, seen face-on (left) and edge-on (right). The top row shows the stellar light, rendered from the star particles in the i, V, and U bands with pynbody. The bottom row shows the surface density of molecular hydrogen ($\rm H_2$) gas on the same scale and orientation. The galaxy has a settled, thin stellar disc, and its $\rm H_2$ gas, where new stars form, is concentrated in the disc plane and along the spiral arms.}
    \label{fig:pix}
\end{figure}

This Milky-Way analog is modeled in its cosmic environment, reaching a virial mass of $6.8\times 10^{11}\, \rm M_{\odot}$ at $z=0$, for a virial radius $R_{vir}=227$ kpc and a stellar mass $M_{*}=4.1\times 10^{10}\, \rm M_{\odot}$. A multi-band optical image and a gas density map are presented in Fig.~\ref{fig:pix}. The spatial force resolution is $\sim 173$ pc and mass resolutions for stellar, gas and DM particles are respectively $m_{*}=8\times 10^{3}\, \rm M_{\odot}$,  $m_{H}=2.7\times 10^{4}\, \rm M_{\odot}$ and  $m_{DM}=1.3\times 10^{5}\, \rm M_{\odot}$. As a result this galaxy is modeled with more than 5 million star particles at $z=0$, limiting the effects of numerical stochasticity on star formation and migration. In addition, this galaxy has not experienced a major merger since $z=3$. The simulation includes prescriptions for UV reionization, cooling (inclusive of metal lines and $H_2$) and re-heating \citep{Shen10}, feedback from SN Ia and SN II supernovae and star formation. 

\subsubsection{Accurate resolution of stellar kinematics.}

A core strength of {\bf h277} for the present analysis is its model for star formation. Unlike other simulations of Milky-Way mass galaxies with similar resolution, {\bf h277} models non equilibrium $H_2$ abundances, hence resolves a multiphase ISM, in particular the dynamically cold phase present at $z<1.5$. Star formation efficiency in this model is a direct function of non-equilibrium $H_2$ abundance followed at the gas particle level, not simply a function of overall gas properties. Star particles are therefore exclusively formed in molecular clouds, which are characterized by smaller velocity dispersions compared to the surrounding gas. This allows star particles to inherit the low dispersion of the gas from which they are born. Hence, {\bf h277} is the first $L_{*}$ simulation to recover populations of young stars born with low-velocity dispersion ($\sigma_{birth}<6\, \rm km.s^{-1}$) as observed in local Milky-Way mass galaxies, along with a solar neighborhood present-day stellar age-velocity relation (AVR) nearly identical to that observed in our Milky-Way \citep{Bird20}. Because the AVR is directly inherited from the progressive radial mixing of old stars (born hotter) and young stars (born colder), this also serves to demonstrate a particularly realistic level of radial mixing in this simulation. It was indeed shown to reproduce with high accuracy the observed decrease of stellar dispersion at birth $\sigma_{birth}$ across cosmic time, especially since $z\sim 1.5$ \citep{Kassin14}. This result is especially important since $\sigma_{birth}$ is known to be strongly correlated to local star formation rates \citep{Kassin14}. While its significance has been studied so far in terms of the vertical component $\sigma_z$, horizontal components of the dispersion are just as important since lower levels of heating can predispose stars to true radial migration at resonance, while disc heating can be a significant contributor to the effective radial migration rates observed in voxel- or pixel-based spatially resolved observations through increase in the orbit eccentricity of stars.

The realistic link between star formation and $H_2$ abundance also tends to concentrate stellar feedback energy into more efficient generation of outflows. These are known to contribute to heating and to trigger a decrease of gas fractions necessary to recover the decrease of $\sigma_{birth}$ over time \citep{Ceverino17}. This results in particular in an appropriate rotation curve \citep{Governato12,Christensen14} and realistic distributions of stellar mass, metallicity and HI.

These characteristics, absent from previous simulations used to quantify the impact of radial mixing on star formation histories \citep{Minchev2026, Bernaldez2026}, set {\bf h277} apart to study the effects of radial mixing on specific spatially-resolved observations of $L_{*}$ galaxies. 

\subsection{Tracking radial migration.}
\label{sec:binning}

Star particles in {\bf h277} have fixed ID numbers that allows for easy tracking across snapshots. In the present study, for each ``mock observation", we quantify radial migration across 4 look-back times $t_{back}=0, \,\sim0.5, \,\sim 1$ and $\sim 1.5$ billion years respectively (For each mock observation, precise lookback times are adjusted to available snapshots).
We track the star particles' radial positions in the frame of the galaxy across time. The frame of the galaxy is centered on the center of mass of the galaxy, with 2 axes (X, Y) within the plane of the disc and the third axis Z taken along the stellar angular momentum $\vec{L}_{gal}$ of the galaxy.

\begin{figure}
    \centering
    \includegraphics[width=1.0\linewidth]{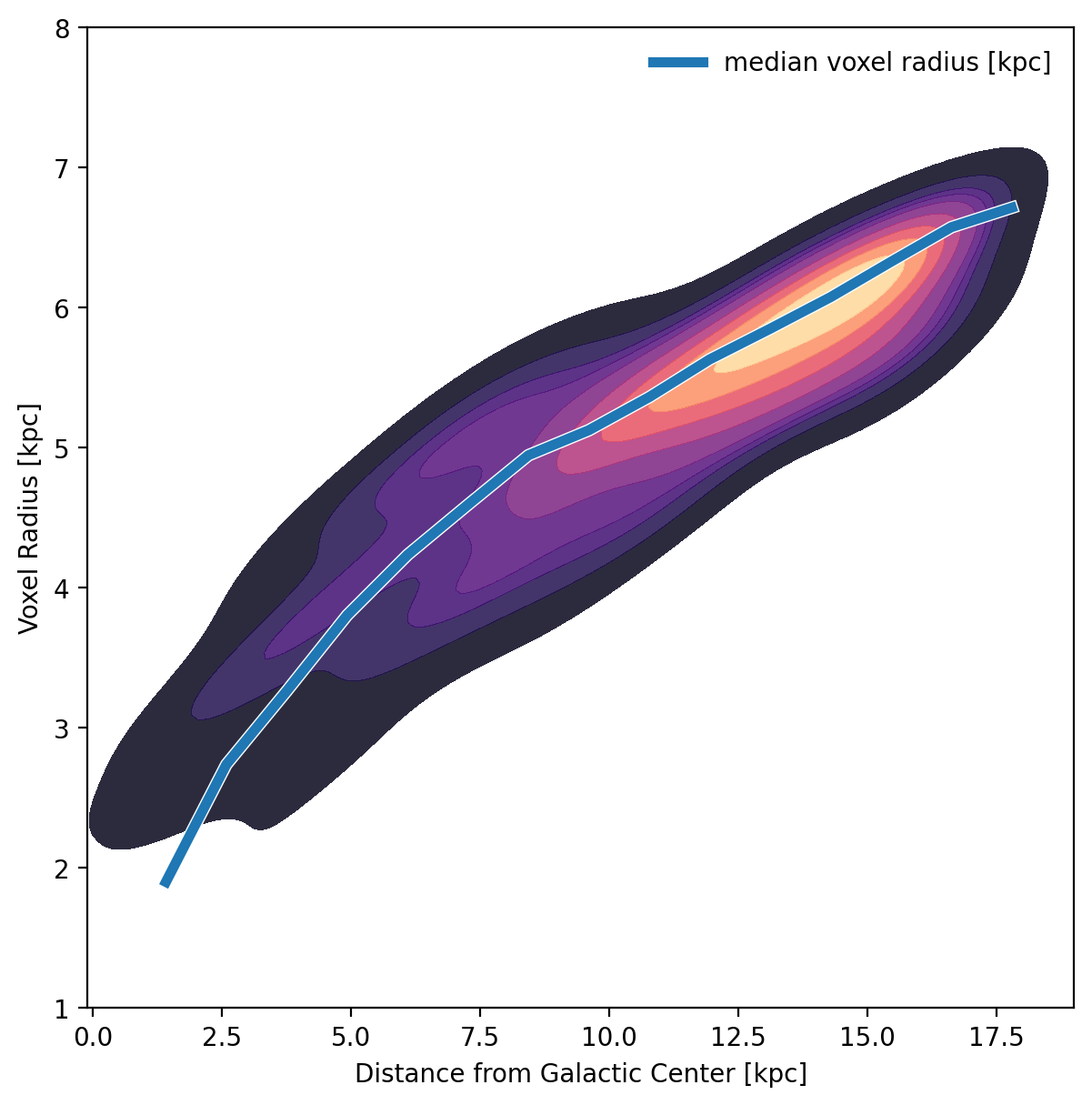}
    \caption{Voxel radius as a function of galactocentric radius for the UVCANDELS galaxies of \citet{Olsen2026}, where each voxel is a Voronoi cell grouping pixels to reach the signal-to-noise needed for multi-band SED fitting. The shading shows the density of all voxels, and the solid line shows the running median. Voxels grow from $\sim2$ kpc in the galaxy core to $\sim8$ kpc in the outskirts, because the fainter outer disc needs larger regions to reach the same signal-to-noise. We use this relation to set the widths of our variable-width radial bins.}
    \label{fig:voxelsize}
\end{figure}

The main objective of our analysis is to move past the study of purely theoretical effects \citep[e.g.][]{Minchev2026, Bernaldez2026} to quantify the actual effects of stellar migration on our ability to recover radial SFR and SFH profiles within the resolution limitations of current surveys. Suitable large statistical samples will be provided by synoptic photometric surveys such as LSST/Rubin. These are therefore our prime targets. In our pathfinder study \citep{Olsen2026}, we specifically demonstrated how to derive such radial profiles from spatially-resolved multi-band photometric SED fitting using a small sample of UVCANDELS galaxies at $z=0-1$.  This showed how, in such photometric surveys, SED fitting methods allowing the reconstruction of SFR and SFH radial profiles typically require the Voronoi re-binning of fluxes in large, high signal-to-noise voxels. Fig.~\ref{fig:voxelsize} shows the median radial extent of such Voronoi cells (voxels) after re-binning across our UVCANDELS sample, taken as representative of this type of reconstruction, as a function of distance from the center of the galaxy. The typical size of a voxel degrades from 2 kpc to 8 kpc as distance to the center of the galaxy increases.

 In the present study, our primary goal is to assess the effects of {\it radial } stellar migration within these resolution constraints, and more generally to assess how radial binning should be adjusted to allow accurate resolution of sSFR radial profiles irrespective of instrumental limitations. We therefore define contiguous annular bins centered on the center of mass of the galaxy. Crucially, the radial width between bins is matched to typical voxel/spaxel size at shell mid-point for face-on spiral type galaxies at $z=0.1 - 1$ in target surveys. In particular, we calibrate the radial shells to the UVCANDELS sample in Fig.~\ref{fig:voxelsize}. We then derive radial profiles using these bins and assess transfers of star particles between them. We also compare our results to profiles obtained for narrower, regular bins more comparable to IFS surveys.

In summary, the following work presents the analysis for two sets of radial shells:
\begin{itemize}
    \item {fine binning:} A regular-width set of circular shells with a circular core of 2 kpc diameter and inter-shell radial width $\Delta r_1= 1 $ kpc, corresponding to the typical angular resolution of spaxels from low-z integral-field spectroscopy surveys like SAMI \citep{Croom12} or Hector \citep{Bryant2024}. This leads to a total of 15 shells.
    \item {variable-width, adaptive binning:} A varying-width set of annular bins with bin width matched to typical voxel size at the same radius in spatially-resolved photometric SED fitting. In particular, we calibrate our current radial shells to spatially-resolved observations of UVCANDELS galaxies at $z=0-1$ studied in \citet{Olsen2026}.   
\end{itemize}
These photometrically-matched bins can then be used to estimate radial profiles of SFR and sSFR at the time of the snapshot but also in the past through the reconstruction of star formation histories in each radial bin.

\begin{figure*}
    \centering
    \includegraphics[width=0.9\linewidth]{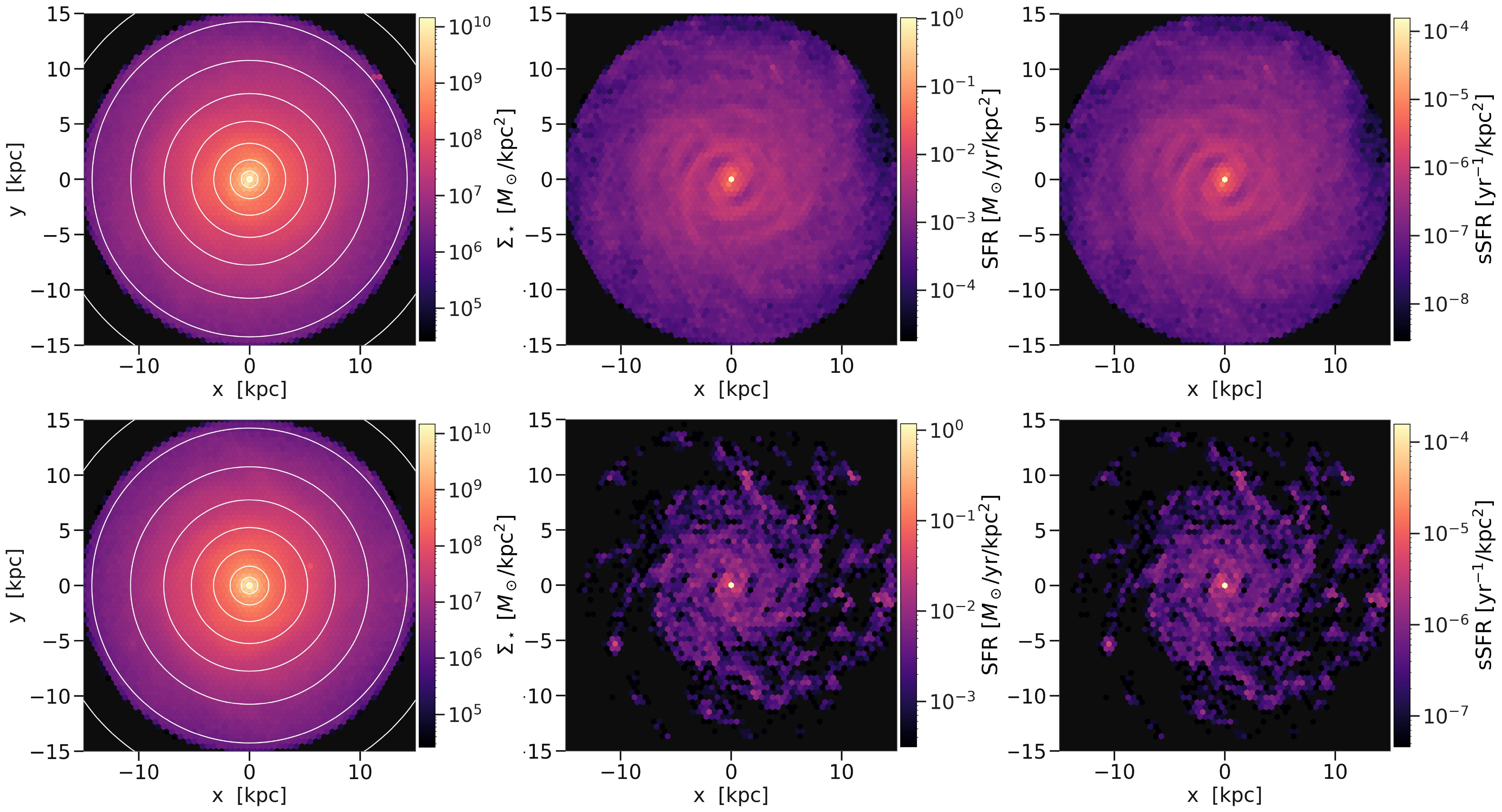}
    \caption{Face-on maps of h277 showing the stellar mass surface density (left), the SFR surface density averaged over 100 Myr (center), and the resulting sSFR (right), in 0.5 kpc pixels. The top row shows the quiescent window at its observation time ($\Delta t=0$, $z=0$), and the bottom row shows the same galaxy at the first epoch, $\Delta t=0.56$ Gyr earlier. White circles on the stellar mass maps mark the edges of the variable-width radial bins listed in Table~\ref{tab:varbins}.}
    \label{fig:snapheatmaps}
\end{figure*}

Fig.~\ref{fig:snapheatmaps} displays maps of stellar surface density, SFR surface density, and sSFR density for the quiescent window and the first epoch at a lookback time of 0.56 Gyr for the simulated galaxy {\bf h277}. Resulting variable-width radial bins are overlaid as white circles on the stellar density map. In every spaxel, SFR and sSFR are simply calculated by sorting the ages of star particles and accounting for cumulative stellar mass formed across a 100 Myr window and total cumulative stellar mass.


\begin{table}
\centering
    \begin{tabular}{||c|c|c||}
    \hline
    Radial Bin (kpc) & $N_{stars}$ & Fraction of total ($\%$)\\
    \hline \hline
    $0<r<1.75$ & 3,100,018 & 44.2 \\
    \hline
    $1.75<r<3.25$ &  1,289,697 &           18.3 \\ \hline
    $3.25<r<5.25$ &  1,098,966 &           15.6 \\ \hline
    
    $5.25<r<7.75$ &   802,952         &  11.4 \\ \hline
    
    $7.75<r<10.75$ &   474,674          & 6.8 \\ \hline
    $10.75<r<15.0$ &    257,917  &         3.7 \\
    \hline 
    \end{tabular}
    \caption{The variable-width radial bins used in this work, for the quiescent window at $\Delta t=0$ (snapshot 00512). For each bin we list its radial range, the number of star particles it contains ($N_{\rm stars}$), and their share of all star particles in the galaxy. The bins widen outward to follow the voxel sizes in Fig.~\ref{fig:voxelsize}. The central bin holds 44\% of the stars, while the outermost bin holds under 4\%.}
\label{tab:varbins}
\end{table}

\subsection{Star formation histories}

In this simulation, the star formation history of each radial bin can be reconstructed two ways:
\begin{itemize}
    \item {\it Ground Truth}: Tracking the galaxy directly across past snapshots corresponding to key lookback times ($\Delta t = 0, 0.5, 1, 1.5$ Gyr), we compute {\bf at each snapshot} the instantaneous SFR and sSFR by calculating the cumulative stellar mass formed in the last 100 Myr directly from the ages of the star particles present in the bin in this snapshot. The interest of directly analyzing past snapshots is that it provides the actual past SFRs without any effect of stellar migration.
    \item {\it SED-like reconstruction}: SFHs in each bin are derived only from the latest snapshot: the ``observation'' snapshot. These are obtained simply by sorting the ages of the star particles in each radial bin in the "observation" snapshot and calculating directly the cumulative stellar mass formed across a 100 Myr window and the total cumulative stellar mass formed for key look-back times ($\Delta t = 0, 0.55, 0.9, 1.55$ Gyr). This method assumes that star particles have not moved from their current radial bin at time of observation (therefore their current bin is assumed to be their birth bin). This measurement corresponds to assumptions commonly made in SED reconstructions from observed data. It is however polluted by radial migration, the effect of which is one of the main uncertainty of the method.
\end{itemize}

We reconstruct the SFHs using the two methods. The comparison allows us to directly quantify the effects of stellar migration on the accuracy of spatially-resolved photometric inference of SFH that neglects stellar migration. Note that we choose 100 Myr under the assumption that the inference of SFR100 from broadband integrated SED fitting should be well constrained with sufficient UV coverage even when SFR cannot be calculated from $H\alpha$ measurements or other SFR tracers. Note that our analysis still provides a ``best case scenario'' where biasing of the SFH is limited to effects of stellar migration between resolution elements and not observational and modeling uncertainties.

\subsection{Evolutionary analysis windows}

\textbf{h277} has a similar SFH to the MW in that it builds much of its stellar mass early. It experiences a series of star bursty major mergers over its first two billion years, undergoes a phase of continuous high star formation with large spiral-associated clumps over the next 2 billions years, then displays steadily declining but still active SF for most of the rest of its history, settling to a constant, relatively quiescent SFR that matches that of the MW by about 3 Gyr lookback time from $z=0$. This demonstrates that h277 is an excellent MW analog. In particular, it is well demonstrated that the \textit{recent} SFH of the MW and h277 at z=0 shows little evolution in SFR from in-situ star formation or recent mergers \citep{Hammer2007, Ruchti2015, Haywood2016, Alzate2021}. However, this steady, relatively quiescent SFR at $z=0$ is not  representative of all disk galaxies and of galaxies at higher z with more stochastic SFHs.

\begin{figure*}
    \centering
    \includegraphics[width=0.95\linewidth]{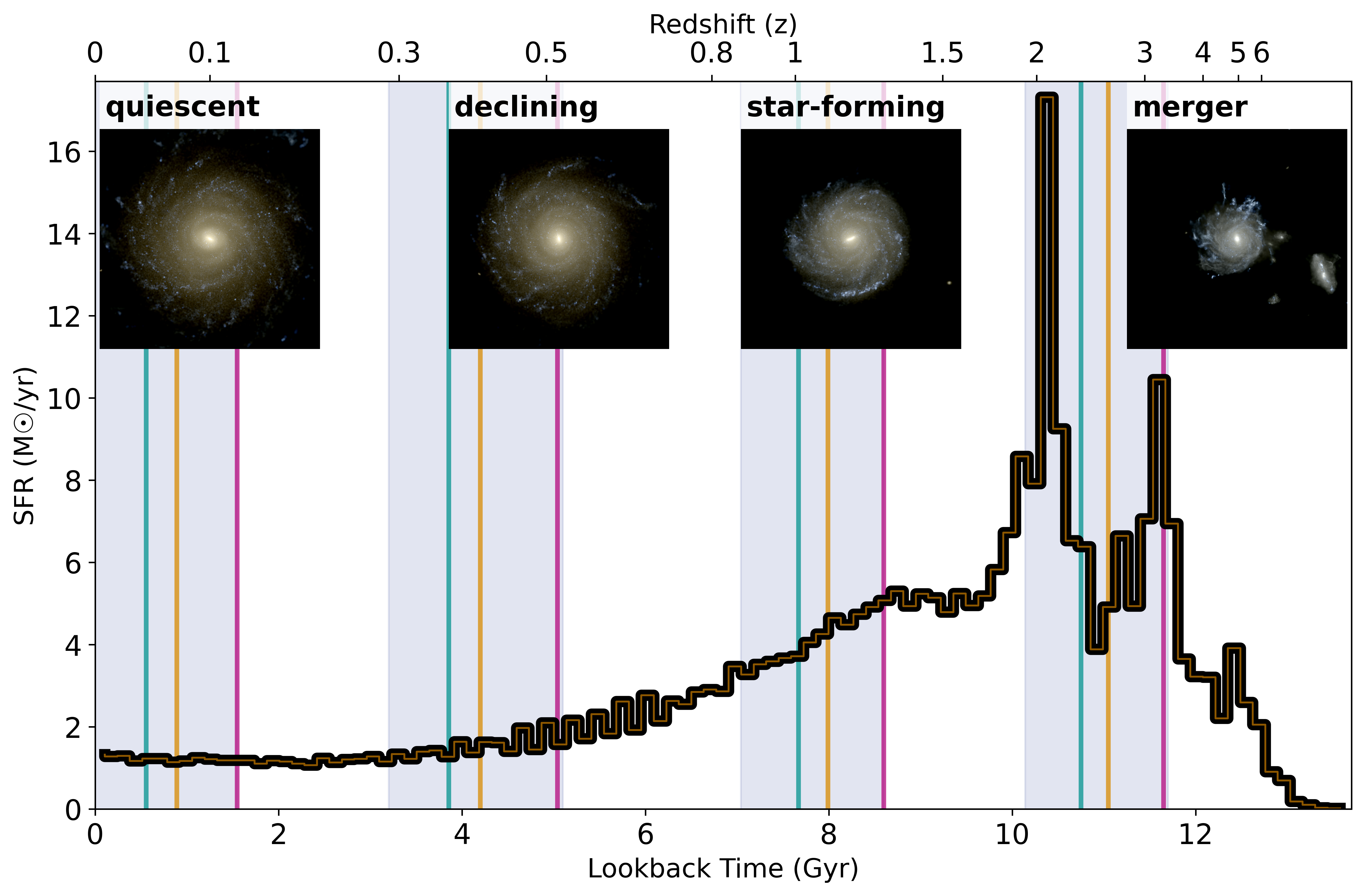}

    \caption{The global star formation rate of h277 as a function of lookback time. Purple shading marks the four analysis windows, each starting at its observation snapshot: quiescent ($z=0$), declining ($z=0.312$, the median redshift of the UVCANDELS sample), star-forming ($z=0.846$) and merger ($z=1.939$). Coloured vertical lines mark the epochs at which we compare inferred and true profiles (Table~\ref{tab:snaps}). The images show the galaxy face-on at each observation snapshot. Together the windows sample a settled, an active but declining, a clumpy, a clumpy and active star-forming and a merging phase of the same Milky Way analog.}

    \label{fig:globalsfh}
\end{figure*}

To explore the quality of inferred spatially-resolved SFHs across a wider variety of star formation scenarios, we therefore reproduce the analysis of h277 over four observation windows. These windows are chosen to represent a variety of evolutionary stages for the galaxy hence be more representative of a variety of galaxies and star formation contexts across redshifts. These windows are shaded in purple in Figure~\ref{fig:globalsfh}. For each window, we define its lower redshift boundary as the "redshift of observation" and we perform both the Ground-Truth and SED-like reconstruction of the SFHs and radial profiles up to $\sim$ 1.5 Gyr in the past(i.e across the purple-shaded window). The specific lookback times used for each reconstruction are highlighted as colored vertical lines. Redshifts of observations include z=0 -- a mostly stable, quiescent phase; z=0.312 -- the median redshift of the UVCANDELS sample and a window of declining but active star formation; z=0.846 -- a window of very active, clumpy star formation (starting at the peak of the steady formation phase), and z=1.939 -- a window covering a known, highly star-bursty major merger. While star formation shows little evolution in the z=0 window, it is more active and stochastic in the z=0.312 window. We anticipate that the time evolving radial profiles from \cite{Olsen2026} may be closest to the z=0.846 window, which is more strongly star-forming. Eventually, the z=1.939 window is highly star-forming with sustained bursts due to a major merger. This window serves as a control for the worst-case scenario in our SFH inference method. We refer to each of these windows throughout the rest of the paper as Quiescent, Declining, Star-forming, and Merger. It is important to note that while the ``Quiescent'' window is relatively quiescent, it is not strictly quenched: the average sSFR  is $7\times10^{-11} yr^{-1}$, as expected for a local low star-forming disc galaxy. It is still the lowest and shows the least evolution in sSFR of all windows. The window description, the relative lookback time for each window, and the global lookback times are shown in Table~\ref{tab:snaps}.

\begin{table}
\centering
    \begin{tabular}{||c|c|c||}
        \hline Evolutionary  &  Relative Lookback & Lookback \\ 
         Window & Time (Gyr) & Time (Gyr) \\ \hline
        Quiescent & 0.0 & 0.0 \\ \hline
         epoch 1 & 0.56 & 0.56 \\
         epoch 2 &  0.89 & 0.89\\
         epoch 3 & 1.55 & 1.55\\ \hline
        Declining & 0.0 & 3.20 \\ \hline
         epoch 1 & 0.65 & 3.86 \\
         epoch 2 & 0.98 & 4.18 \\
         epoch 3 & 1.8 & 5.04 \\ \hline
        Star-forming & 0.0 & 7.05 \\ \hline
         epoch 1 & 0.63 & 7.68 \\
         epoch 2 & 0.94 & 7.99 \\
         epoch 3 & 1.56 & 8.61 \\ \hline
        Merger & 0.0 & 10.14 \\ \hline
         epoch 1 & 0.6 & 10.75 \\
         epoch 2 & 0.91 & 11.05 \\
         epoch 3 & 1.51 & 11.66 \\ \hline

    \end{tabular}
    \caption{The snapshots used in each evolutionary window shown in Fig.~\ref{fig:globalsfh}. For each window we list the base (observation) snapshot and three earlier epochs. The relative lookback time is measured from the window's base snapshot, and the absolute lookback time from $z=0$. Each epoch is the available snapshot closest to 0.5, 1 and 1.5 Gyr before the base snapshot.}
    \label{tab:snaps}
\end{table}

\section{Recovery of radial SFH profiles }
\label{sec:radial_profiles}

In the following section, we assess the quality and accuracy of radial sSFR profiles one can expect to recover from variable-width binning SED inference of spatially-resolved star formation histories by performing a systematic comparison with the ``ground-truth" profiles.

\begin{figure*}
    \centering
    \includegraphics[width=0.95\linewidth]{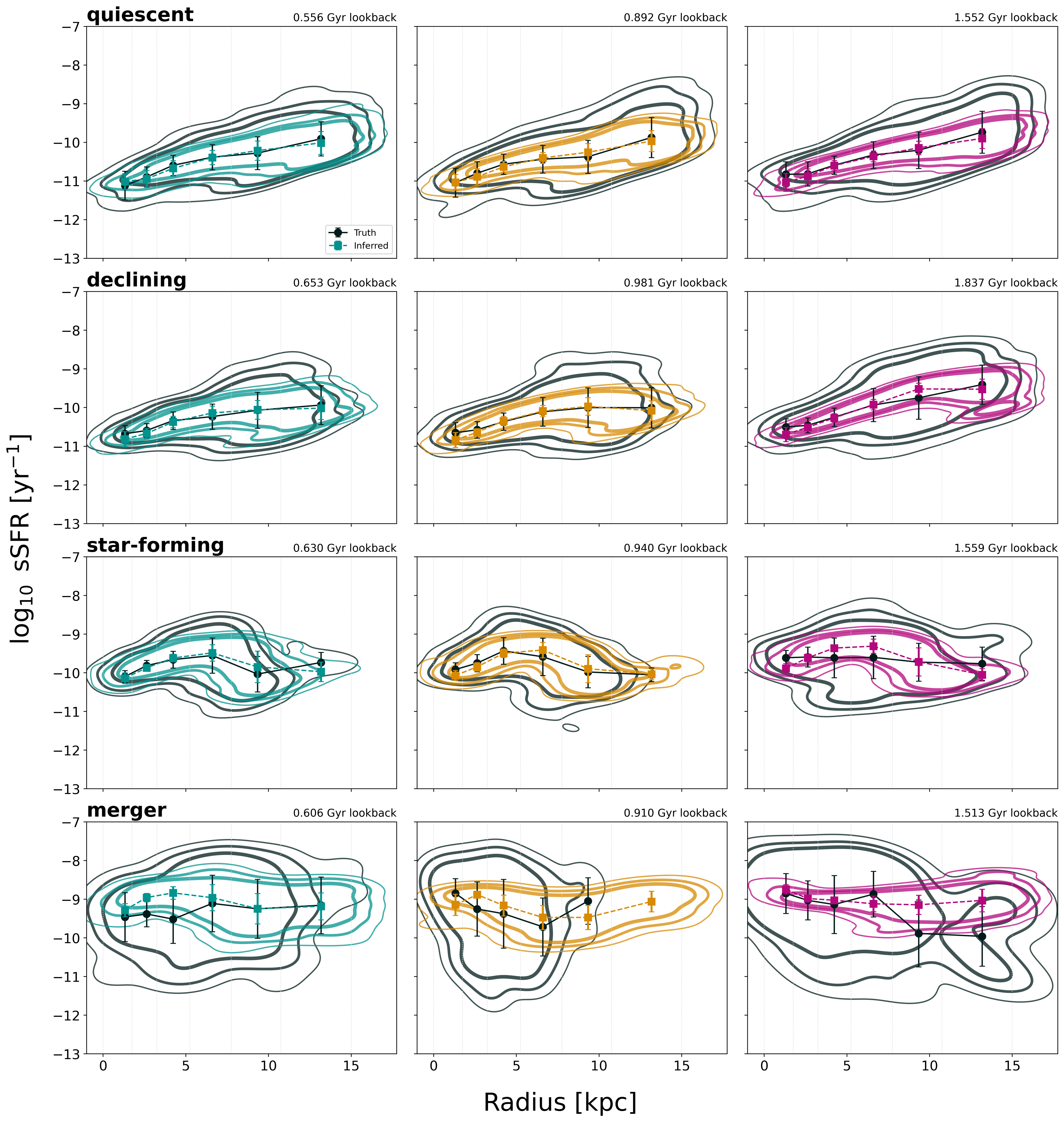}
   \caption{Radial sSFR profiles in variable-width bins for the four evolutionary windows (rows), compared at each epoch. Black shows the true profile measured directly in the snapshot at that epoch. Colors show the profile inferred from the star formation histories of the observation snapshot: teal for $\Delta t\approx0.6$ Gyr, orange for $\approx0.9$ Gyr and red for $\approx1.6$ Gyr. Contours show the distribution of individual spaxels, solid lines show the bin medians, and pale gray vertical lines mark the bin edges. Outside the merger window, the inferred median profile matches the truth at every epoch, but the inferred scatter is systematically narrower because radial migration smooths out past bursts of star formation.}
    \label{fig:radialprofiles}
\end{figure*}

Figure~\ref{fig:radialprofiles} displays the median radial profiles of sSFR and the spaxel-based radial distributions of sSFR (contours) in our four evolutionary windows, from quiescent to merging. Colored profiles are inferred at lookback times of $\Delta t\sim 0.6,$ $0.9$, and $1.6$ Gyr (in teal, orange, and red contours, respectively) and compared to ground truth profiles (in solid black contours).

It is noticeable that, for all windows except the merger case, the inference across photometric bins accurately recovers the median binned radial profiles despite radial migration. Note that this also generally holds for inference across regular 1 kpc-wide (IFU-like) bins presented in the Appendix.~\ref{appendix:finerp}, although in this latter case the bin-per-bin accuracy is already visibly reduced in the outskirts of the galaxy as well as in particularly star-forming bins (see for instance row 3 of Fig.~\ref{fig:fineradialprofiles}). It is not unexpected as the 1 kpc-width of IFU-like shells is reaching the scale of large, fast migrating star-forming clumps, which are therefore not accurately tracked by the static inference.  We present the characteristics of each phase in further details in the next section.

Although median profiles recovery is accurate, it is striking across all windows that the scatter of the underlying sSFR distribution is systematically reduced in the inferred radial profiles compared to the truth values. This result holds true across the full radial range probed here. This is of course an expected effect of stellar mixing across spaxels, which tends to smooth out sSFR values across time. However, let us stress that this general offset does not necessary preclude scatter analysis as relative radial variations might still be preserved. We discuss the quantitative analysis of slope and scatter in Section.~\ref{sec:quantitative}.

\subsection{sSFR profile quality across evolutionary phases}

\subsubsection{Quiescent phase}

Unsurprisingly, the quiescent phase (top row of Fig.~\ref{fig:radialprofiles}) provides the best match between ground truth and inferred profile. Throughout the window, the radial extent of the distribution is recovered within $<6\%$ error. The median trend is matched with less than $1\%$ mean-square error (MSE). 
It is visible that the strongest deviations between inferred and true profile (although still limited to a few percents) are found in the central part and the extreme outskirts of the galaxy, and most pronounced for the largest lookback time ($\Delta t\sim$ 1.5 Gyr). On average, the quiescent phase has an average 44$\%$ error in scatter, but this is a relatively flat systematic that is consistent across all evolutionary stages except the merger window. 


\subsubsection{Declining star formation phase}

The median profile and radial extent is well recovered again (within a few $\%$ error) in this epoch. But qualitatively it is notable that the inferred profile fails to fully recover the sharp increase in spread in sSFR outward of $7$ kpc from the center . Therefore, the inferred profile does not fully account for the coexistence of more star-forming regions with more quiescent ones in the same radial bin. Indeed, comparing the actual spread with the image seen in Figure~\ref{fig:globalsfh}, one can see bluer star-forming regions loosely tracing the spiral arms in this radial range. The much narrower radial spread in sSFR inferred from the SFHs demonstrates that while the median value of the sSFR can be recovered easily past 1 Gyr lookback time, small-scale, spiral associated clumpy star formation is in part smoothed by stellar mixing in the SFHs based profile reconstruction. Note however that inferred radial profiles do see a sharp, albeit low amplitude increase in spread, but it is shifted at a higher radius ( $r\sim 10$ kpc) compared to real profiles. This shows that, despite the smoothing effect, the existence of jumps in sSFR spreads is in fact detectable from the inference, but their amplitude underestimated and their location only accurate to a few kpc precision.

\subsubsection{Active star-forming phase}

From Figure~\ref{fig:globalsfh} the active star-forming phase shows well defined star-forming spiral arms with large bright star forming clumps (kpc-scale) near $r = 7.5$ kpc at time of observation. Both the inferred and actual radial profiles show a peak in sSFR caused by this enhanced star formation in the spiral arms at all snapshots. In fact, this bursts of star formation mediated by kpc-size clumps result in a non-monotonous median radial profile of sSFR, which both the SFH inference recover at $<10\%$ precision out to 1 Gyr lookback time. Note, however, that, like in previous windows, the spread and overall stochasticity across the disk remain underestimated for the distributions inferred from the SFHs, while the sharp variations in relative spread can be shifted by up to 2 kpc compared to ground truth. This still ensures that star-bursty star formation mediated by large clumps is detectable through SFH inference, but stellar mixing induces a lack of precision in its precise location.

What is more, the large lookback time epoch ($\Delta t \sim 1.5$ Gyr) starts to display stronger deviations in both median trend and spread. This justifies limiting the lookback time analysis to $\Delta t\sim$ 1 Gyr for such clump dominated, strongly star forming galaxies.

\subsubsection{Merger}

As expected, the merger window reaches the limits of spatially-resolved SFH inference methods. Beyond the mere average sSFR across the entire galaxy, the inferred profiles fail to recover any spatial information on star formation activity: the spatial extent of star formation can reach a $100\%$ error at $\Delta t \sim 1$ Gyr, and the median radial trend also displays large errors. Similarly, the inference fails to recover both absolute and relative variations in spread, even at the qualitative level. In fact, we see that the inferred profiles capture neither the shape nor the spread of the sSFR distribution in the actual snapshots.

This is an expected result, as the location and distribution of star particles varies widely across snapshots in the window, as progenitor galaxies are violently relaxing into the new remnant disc \citep[for description of violent relaxation,see][]{Lynden-Bell67}. With our methodology, we therefore cannot expect to anticipate the dynamics of a merger characterized by wide-ranging, fast variations in potential and mass distribution. While it is expected, this highlights the necessity for careful selection of galaxies that do not show signs of having experienced a recent major merger (tidal features, counter-rotating cores, higher order kinematic signature) before attempting a SFH based reconstruction of sSFR profiles.

\subsection{Slope, scatter, and mean recovered sSFR of radial profiles}

In the following section, we quantify the overall performance of SFH based inferences in their ability to recover the key properties of radial sSFR profiles: fitted slope, bin-wise sSFR scatter, and the average bin-wise sSFR across all tested evolutionary windows. We further compare the two radial binning schemes we defined in section~\ref{sec:binning}: variable-width radial bins typical of voxel-based photometric reconstructions and fine kpc-wide regular binning achievable with modern integral-field spectroscopy. 

In the context of sSFR radial profiles, agreement in \textit{slope} is an important metric to evaluate the recovery of overall star-formation gradients. Such gradients are key to discriminate inside-out and outside-in star-formation/quenching, which point to different physical mechanisms. On the other hand, the width of the scatter shows how well the SFR reconstruction can capture the level of stochasticity in star formation across the disk. Perfect agreement of this metric with ground-truth values would point to perfect inference of the location and timing of starbursts throughout the disk. While this scenario is not realistic due to the blurring effects of stellar migration, relative radial variations can already be informative on the approximate location of large star-forming clumps. 

\label{sec:quantitative}
  \begin{figure*}
    \centering
    \includegraphics[width=0.95\linewidth]{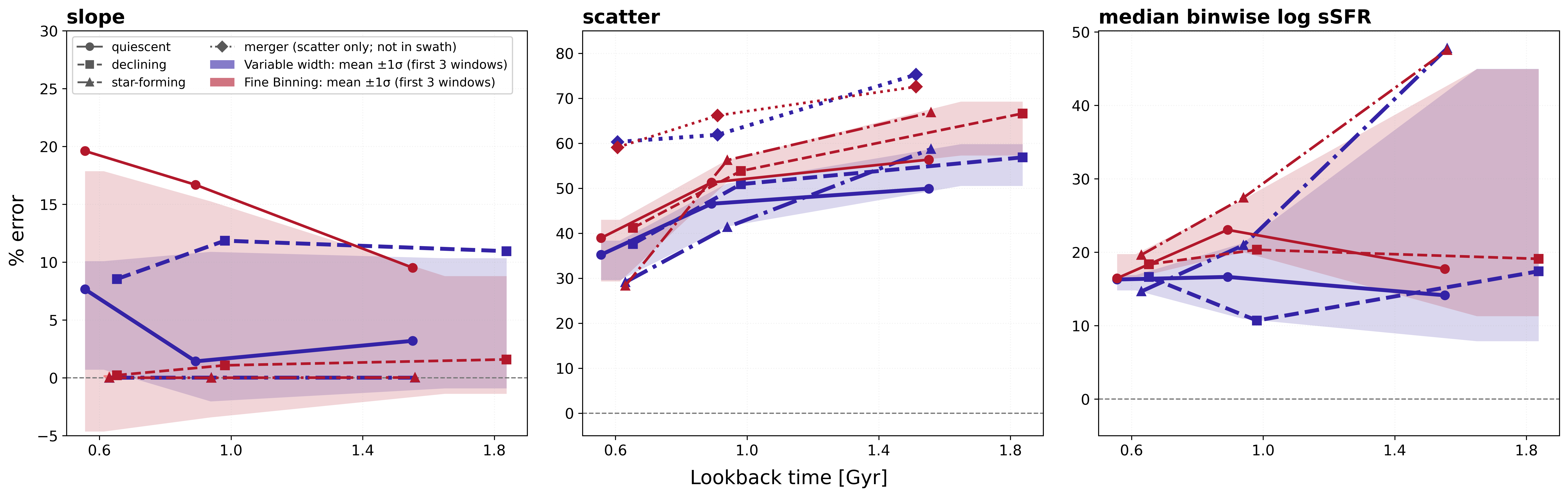}
    \caption{Percent error between the inferred and true radial sSFR profiles as a function of lookback time, for the slope of the profile (left), the scatter within bins (center) and the bin-median log sSFR (right). Each window is shown for variable-width bins (blue) and fine 1 kpc bins (red). The shaded bands show $\pm1\sigma$ about the mean of the quiescent, declining and star-forming windows. The merger window is shown only in the center panel, because its slope and median errors are far larger than the rest. Both binning schemes recover the slope to within 20\%, but variable-width bins give smaller errors in both the scatter and the median sSFR.}
    \label{fig:slopescatter4windows}
\end{figure*}

 Figure~\ref{fig:slopescatter4windows} shows relative errors between inference and ground-truth for the variable-width binning in blue and for the fine regular binning in red. The percent difference in slope and mean sSFR for the merger window  is omitted for visibility as it is far greater than in other windows and we have already established the poor predictive power of SFH-based reconstructions in this regime. It is however included in the central panel comparing scatter. The merger window remains an outlier regardless of binning scheme.
 
 To highlight the overall trends for each binning scheme, we include a shaded region of 1$\sigma$ around the mean of the quiescent, declining, and star-forming windows. Note that the star-forming window is compatible with a slope of 0 across methods and binning schemes. To avoid artificially large relative errors, fitted slopes compatible with 0 at the bootstrap level 0.5$\sigma$ (all below observational detectability levels) are assigned a value of 0.

The degree of recovery of the slope depends on the selected temporal window, epoch, and binning strategy; however, the relative error remains below $20\%$ for all conditions evaluated. The greatest discrepancy is observed in the quiescent window, in which the variable-width binning outperforms the fine binning by approximately 10–15\%. Fine binning exhibits reduced accuracy at earlier epochs, with a $\sim$20\% error at $\Delta t \sim 0.6$ Gyr that decreases progressively to $\sim$10\% by epoch 3. In contrast, the fine-binning method recovers the slope of the declining window at higher accuracy than the variable bin-width scheme: deviations remain under 3\% in all epochs, while the variable bin setup shows offsets of roughly 8-12\%. This is mostly owing to the very smooth and plateauing radial variations in slope, which mitigates bias in the outskirts, otherwise common with fine binning. The resulting differences in the estimated slope therefore likely stem from a slight underestimation of sSFR in the galaxy’s central region that is slightly more pronounced for variable-width binning. Indeed, while this bias appears in both binning approaches, the denser sampling provided by fine binning yields a closer agreement with the reference slope in this range. In the star-forming window, fine and variable binning give similar results; however, this similarity mainly arises because the slopes in that window are all compatible with zero. We provide higher-order analysis of these non-monotonous profiles in Appendix.~\ref{fig:peakoffset}.

Variable binning still displays better overall performance in slope recovery as the error remains strictly under $12\%$ across lookback times, evolutionary phases and radial gradient behaviors, While fine binning can reach $20\%$ error in conditions where the radial gradient remains high in the outskirts, where this binning shows larger biases.

For both binning prescriptions, the scatter is systematically underestimated by \(\gtrsim 30\%\), with the magnitude of the bias moderately increasing with lookback time. Nevertheless,at all epochs the variable-width binning scheme yields a systematic improvement relative to fine binning, typically reducing the scatter error by 5 to 15 point percents (amounting to $\sim 10\%-30\%$ of the total scatter). As a result, for all windows but the merger window, variable binning maintains an error on the scatter below $52\%$ even at $\Delta t\sim1.5$ Gyr lookback time.
Note that for the merger window, the percent error on the scatter jumps to  \(\sim 60\%-75\%\) across epochs even for the variable binning scheme. This confirms the low reliability of SFH-based inference methods in this regime.

That variable-width binning provides a measurably more accurate estimate of the scatter than fine binning indicates that the choice of spaxel size and the radial binning strategy can partially mitigate the smoothing and blurring introduced by radial migration in the presence of bursty star formation. Although the scatter discrepancy increases with epoch, the variable-width scheme remains at the \(\sim 52\%\) level across windows, whereas fine binning by epoch 3 is comparable to, or worse than, the merger-window discrepancy at the first two epochs, approaching \(\sim 65\%\). We therefore conclude that scatter will be underestimated irrespective of binning approach; however, to recover physically informative constraints on the radial locations of starbursts beyond \(\sim 500\) Myr, observers should rebin the data using radial bins that increase in width with galacto-centric radius so as to mitigate inter-bin stellar migration.

For both binning prescriptions, the bin-wise median log sSFR is generally well recovered for all windows within $\Delta t=1$ Gyr lookback time and even $\Delta t\sim1.5$ Gyr for the quesicent and declining phases. However the percent error increases in the high-star forming window past $\Delta t=1$ Gyr. Overall, we measure a systematic error  $<10-17\%$ across most windows and lookback times (apart from the high star forming window at $\Delta t\sim1.5$ Gyr) for the variable width, a marked improvement from fine regular binning for which errors are almost systematically $>20\%$. 

As mentioned however, the median bin-wise log sSFR  error reaches $45\%$ and $55\%$ at the largest lookback time under the variable-width and fine-binning methods respectively. This is not unexpected as Fig.~\ref{fig:radialprofiles} shows that, in this phase, the star formation is mediated by large spiral-associated clumps (appearing as peaks in the radial profiles), the location of which is only precise at the 1-2 kpc level due to faster migration through the disc. This offsets the precision on sSFR across several bins. Hence, even with adaptive radial binning, systems undergoing vigorous star formation warrant additional caution, as SFH-based inferences produce stronger bin-level sSFR biases in this regime. This bias is however consistently more severe when fine regular bins are adopted.  

To limit the overall mean-sSFR underestimation to $<$20\% across evolutionary windows, we recommend adopting variable-width binning. For actively star-forming galaxies with non-monotonous sSFR radial profiles, SFH reconstructions should be considered reliable only out to a lookback time of $\sim$1 Gyr, beyond which biases become substantial.

\subsection{Relative scatter variations across sSFR profiles}

\begin{figure*}
    \centering
    \includegraphics[width=0.95\linewidth]{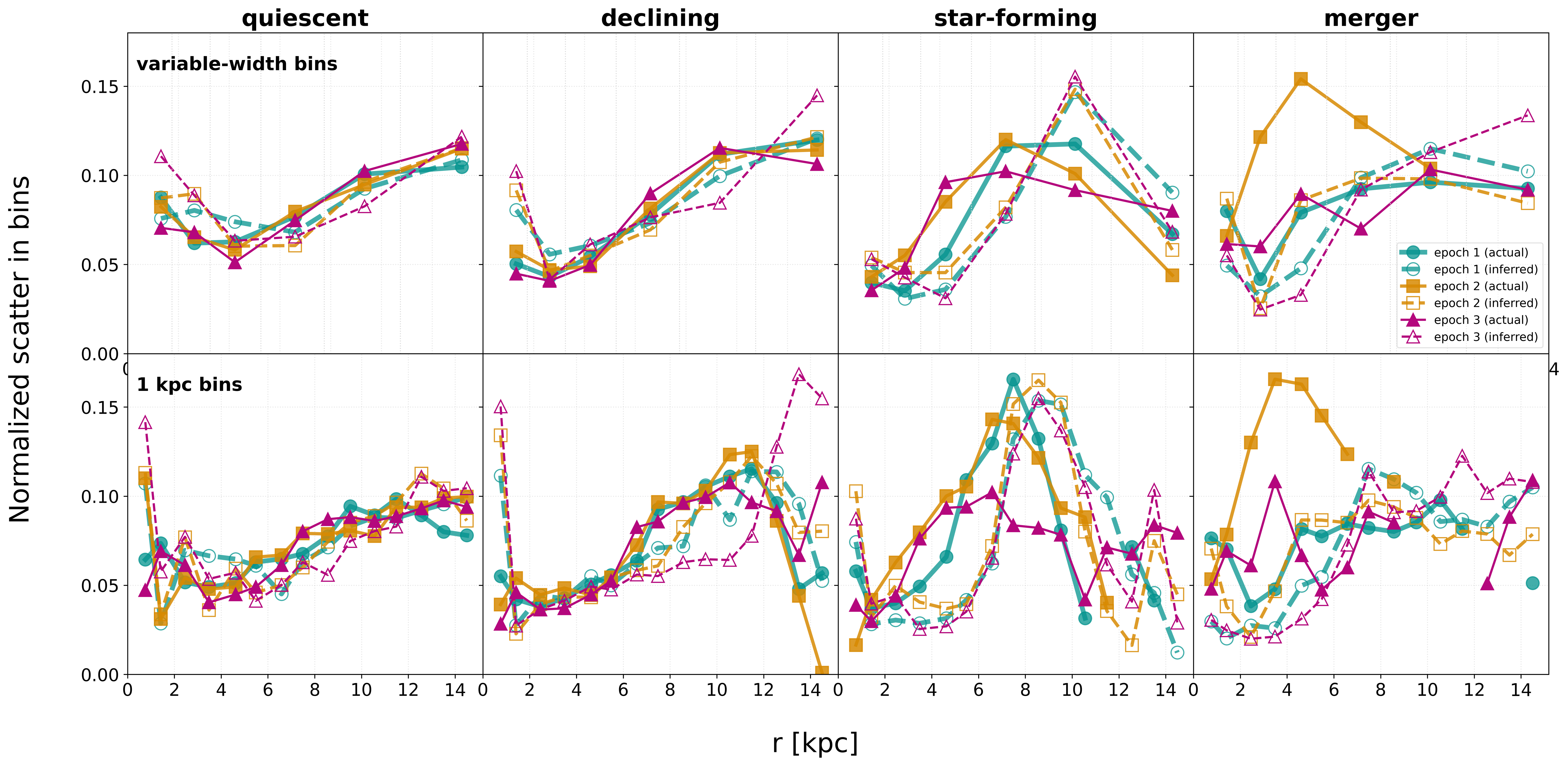}
    \caption{Normalized scatter $\sigma_{norm}$ (Eq.~\ref{eqn:signorm}) as a function of galactocentric radius for each window (columns), in variable-width bins (top) and fine 1 kpc bins (bottom). Normalizing removes the overall underestimate of the scatter, so the panels compare only its radial shape. Solid lines show the truth and dashed lines the inferred profiles, with colors marking the epochs. Outside the merger window, the inferred scatter keeps the radial shape of the true scatter, and variable-width bins match it best in the center and outskirts. In the star-forming window the peak of the scatter is shifted outward by 1--2.5 kpc (Fig.~\ref{fig:peakoffset_scatter}), and in the merger window the shape is not recovered.}
    \label{fig:scatterperbin}
\end{figure*}

Figure~\ref{fig:slopescatter4windows} confirmed the overall systematic in scatter, whereby the SFHs tend to smooth over bursty star formation and bring overall sSFR closer to the median at all epochs. But such a systematic effect across bins does not necessarily preclude the analysis of relative variations with radius. In particular, while the spread in sSFR may be partially smoothed due to radial migration, is the radial location of outstanding star-forming features in the disk preserved, even as the true level of sSFR stochasticity is not? To further explore this, we calculate the normalized fractional scatter in each bin index by its median radius $r_0$ as : 

\begin{equation}
    \sigma_{norm}(r_0) = \frac{\sigma(r_0)}{\int^{r=r^{bin}_{max}}_{r=0} \sigma(r) dr}
    \label{eqn:signorm}
\end{equation}

for both inferred and actual curves. In Figure~\ref{fig:scatterperbin} we plot the normalized scatter of the actual distributions (solid lines) with the inferred distributions (dashed lines) for each evolutionary window for variable width bins (top) and fine bins (bottom). 

For the first two windows, characterized by mostly smooth, monotonous evolution across the disc outside of the central bins, the overall trend is consistent between inferred and actual profiles across all radial bins. This consistency suggests that the radial information encoded in scatter variations is retained, even when the absolute sSFR spread of substructures within the disk is underestimated. The inference based methods are able to recover both the peak of sSFR spread in the center and the smooth increase towards the outskirts across the disc. it is notable that the variable-width binning performs better than fine binning overall. This higher performance is driven mostly by its increased accuracy in the most central parts and extreme outskirts of the galaxy, where stellar migration is more important and starts to strongly bias fine binning through inter-bin migration. Unsurprisingly, this bias becomes more pronounced at higher lookback times (pink lines).

In the star-forming window (third panel), deviations become more pronounced. The variable binning scheme still allows the reproduce the central values and to detect the existence of a peak in sSFR scatter at $r\sim 6 -10$ kpc from the center of the galaxy (associated with clumpy spiral arms), but its position is shifted compared to ground-truth. The shift is limited to $\sim 1$ kpc at $\Delta t\sim 0.5 $ Gyr but steadily increases to $\sim 4-5$ kpc at $\Delta t\sim 1.5 $ Gyr (see Appendix.~\ref{apendix:offset} for more details on shifts in sSFR and sSFR spread, with proper gaussian fitting of the profiles and quantification of inter-peak separations). This provides more incentive to limit inference of radial profiles to $z<1$ galaxies and to $\Delta t\sim 1 $ Gyr lookback times, i.e below the galaxy's typical dynamical time, in detectably clumpy, highly star forming cases like window 3.
These shifts are also present in the fine binning scheme as they are a consequence of the inability of SFH-based inferences to properly account for relatively fast processes like resonance-based stellar mixing and migration of large clumps through the disc. Despite these spatial offsets, fine binning reproduces the relative level of stochasticity of the clumpy mid-disc region more accurately, at least at short lookback times $\Delta t <1$ Gyr, owing to its increases radial sampling. However, once again, this comes at the cost of much reduced accuracy in all other areas, in particular in the central parts and outskirts of the galaxy, where variable binning provides a much better fit.


Results for the merger window are presented for reference on the fourth panel but only confirm our previous diagnostic: SFH-based reconstructions fail to recover even relative variations of spread.
In conclusion, our analysis shows that SFH-based inference are remarkably efficient at recovering smooth variations in sSFR scatter across the disc out to 1.5 Gyr lookback time and reasonably locate spread variations due to star forming clumps out to 1 Gyr in the past, but fail to spatially track the build-up of said large, highly star-forming star forming clumps or accurately locate the variations due to their migration.

\section{Radial migration across 1.5 Gyr}
\label{sec:migration}

The accuracy of radial sSFR profiles obtained through inference across 1.5 Gyrs and across a variety of star formation patterns is remarkable. In particular, the enhanced overall performance of varying ``photometric bins" compared to IFU-like regular, narrower radial bins shows that optimal radial binning is strongly underpinning this accuracy. Indeed, while significant amounts of stellar migration are expected across 1.5 Gyr of evolution \citep{Minchev2026}, its effects on resulting sSFR profiles are bound to depend on:
\begin{itemize}[leftmargin=*]
    \item the mean radial distance traveled by star particles born in any given radial bin, which determines the fraction of star particles that transfer between bins over cosmic time.
    \item the balance between inbound and outbound star particles in any bin, which regulates whether the resulting inferred profile will represent either a mostly smoothed or a highly distorted version of the true profile.
\end{itemize}

\begin{figure*}
    \centering
    \includegraphics[page = 1, width=0.95\linewidth]{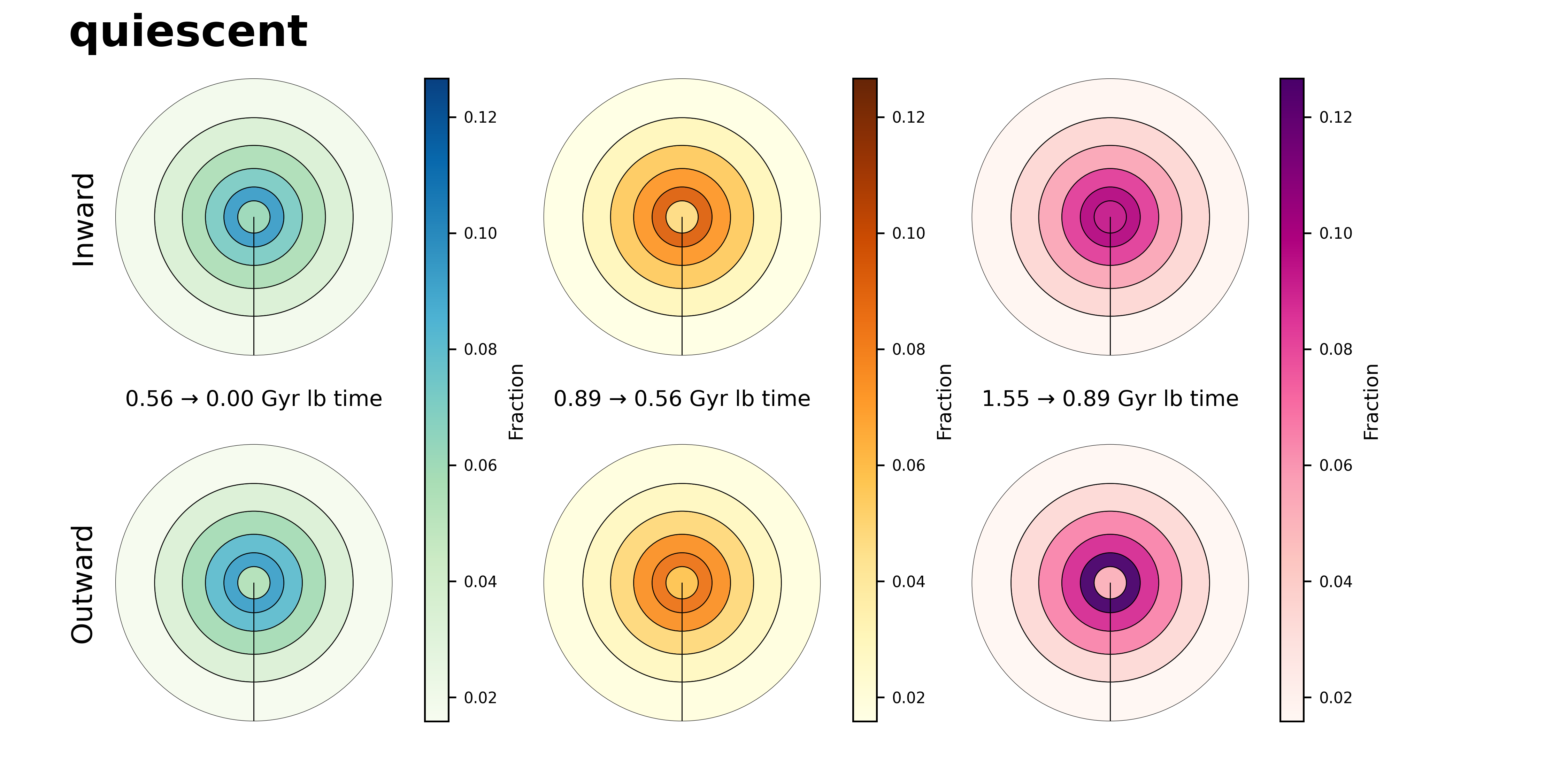}
    \includegraphics[page = 2, width=0.95\linewidth]{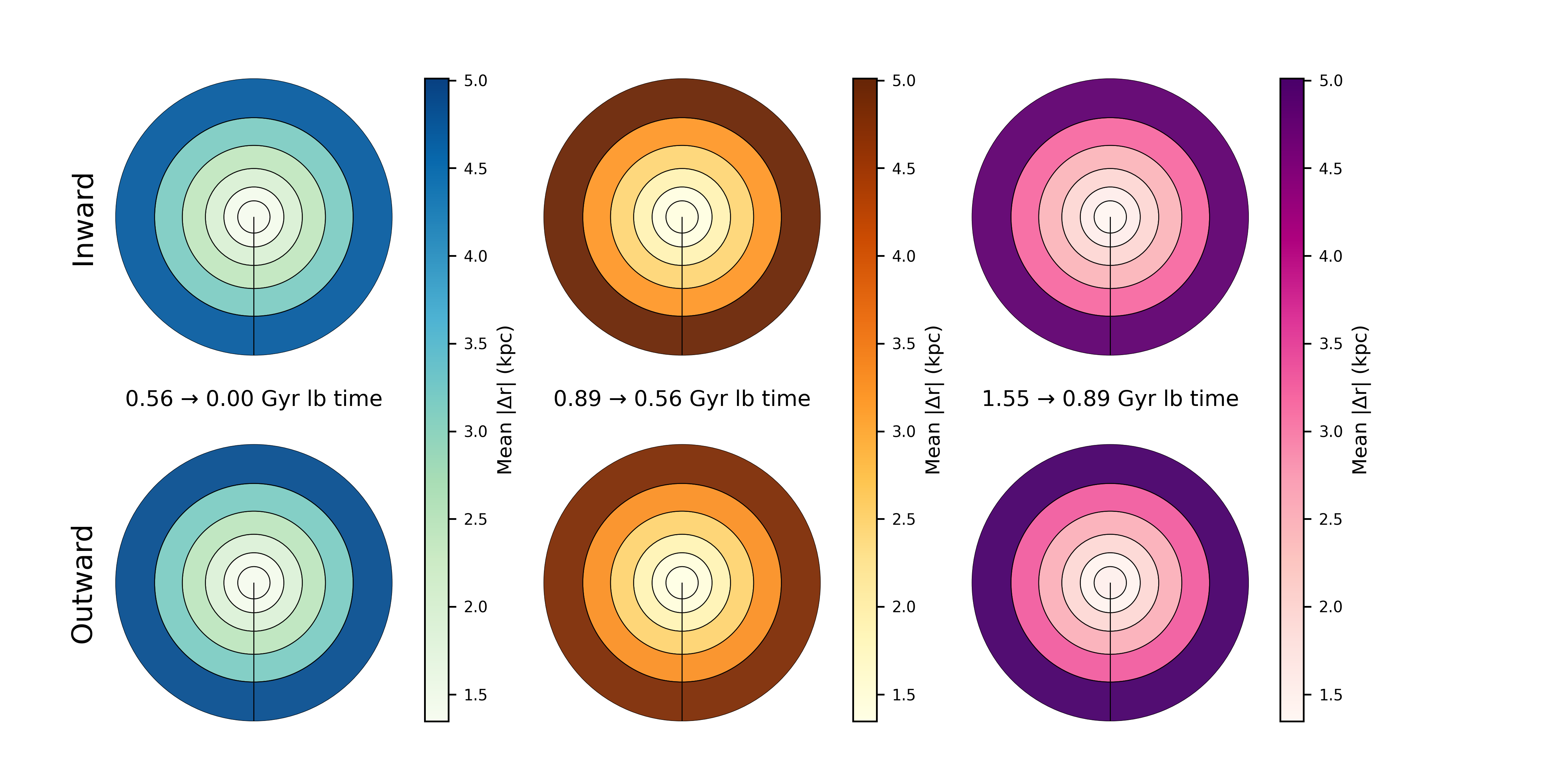}
    \caption{Bin-to-bin migration in the quiescent window with variable-width bins, for each $\sim0.5$ Gyr step between epochs (columns). In the top figure, each ring is a radial bin colored by the fraction of all star particles in the galaxy that moved into it (Inward row) or out of it (Outward row) during that step. In the bottom figure, the same bins are colored by the mean radial distance traveled by those bin-crossing particles. Bin crossing is most common near the center, but these particles move only $\sim1.4$ kpc on average, less than the width of the central bin, and inward and outward movement nearly balance. The few particles that cross the outermost bin travel $\sim5$ kpc, which the widest bins absorb, supporting bin widths that grow with radius.}
    \label{fig:window1_migration}
\end{figure*}

To optimize guidelines for binning strategies, we now examine the amount of stellar mixing across our choice of radial bins, by tracking the position of star particles across each epoch. Figure~\ref{fig:window1_migration} focuses on the quiescent window and displays maps of the inward and outward migration fractions across bins during successive 0.5 Gyr timesteps (top two rows), and maps of the mean distance travelled by bin-crossing stars, inward and outward (two bottom rows). Similar figures for the other three windows are shown in Appendix~\ref{appendix:migration}, but a major common trend is that the amount of migration roughly scales with the width of the bin, which in turn limits the fractions in and out of each bin. The result is that the fractions of star particles that have migrated remain low ($<4\%$) everywhere except the center (where it remains $<13\%$). Comparing this with the mean distances travelled by displaced stars, we see that the mean shift in the center is less than the width of the innermost bin at $1.4$kpc, which indicates that the impact on the radial profiles would be minimal. 

The outskirts have very few star particles that cross in and out of their bins, but the mean displacement for these can be up to 5 kpc. Overall, this large mean shift has little impact since so few star particles move out of these larger bins.

This remains the case for the other evolutionary windows though, in the active star-forming window presented in Fig.~\ref{fig:window3_migration_bin_dist_compare} (and Figure~\ref{fig:window3_migration} in Appendix.~\ref{appendix:migration} for migration rates), there is a larger fraction of cross-bin migrating stars within $6-8$ kpc from the center reflecting the faster migrating clumps associated to spiral arms and resonances. However, even at greater lookback times, cross-bin transfer fractions remain $<16\%$ and the mean traveled distance over 0.5 Gyr below $1-2$ kpc for variable binning. This explains why the effect of stellar migration on reconstructed profiles in this window is limited to a spatial offset in sSFR and sSFR scatter peaks of that order of magnitude.   

Results for the merger window are discussed in Appendix~\ref{appendix:migration}. These confirm that violent relaxation triggers a dramatic increase in both the fraction of star particles migrating out of any given bin (reaching up to $60\%$) and the mean distance traveled by bin-crossing stars (reaching up to 13 kpc, hence moving across the face of the galaxy). Overall, our results across all windows show that our choice of radial binning works exceptionally well in cases when the disk is relatively settled and follows a steady accretion model but breaks down in the presence of a major merger. 

It should be noted that the fine-binning scheme reproduces the global trends in mean radial migration distances observed in the variable-binning configuration but this now means that many stars travel several kpc-wide bins within 0.5 Gyr and fractions of crossing stars are, unsurprisingly, much increased . Fig.~\ref{fig:window3_migration_bin_dist_compare}, bottom rows show migration metrics with fine binning for the star forming window, revealing in much more detail the trajectory of sling-shot clumps near resonance, while Appendix~\ref{appendix:migration} presents results for the quiescent window shown in Figure~\ref{fig:window1_migration_1kpc}. Examination of the mean distances traveled into and out of the radial bins indicates that the bins exhibiting stellar displacements in excess of twice the bin size now 
extend $\sim1.4–2.3$ kpc inside 5.25 kpc, rising to $\sim4–6$ kpc at 5–11 kpc around the center of the galaxy. This explains why adopting wider bins in this radial range, with bin extent sufficiently in excess of typical distance traveled by migrating stars, yields closer agreement in the actual scatter.

Note also that apart from reproducing the same radial gradients observed in Figure~\ref{fig:window1_migration_1kpc}, the fixed, fine radial binning yields limited additional insight relative to the variable-width binning, with the primary exception of the amplitude of scatter in the  high scatter regions, clumpy regions of the highly star forming window.
Fig.~\ref{fig:window3_migration_bin_dist_compare}, bottom row, allows to appreciate how fast these larger substructures migrate across bins compared to diffuse stars. Average mean radial distance traveled can reach $11$ kpc in single fine bins, hinting at at least some structures being on highly eccentric orbits and confirming that, irrespective of binning, SFH-based inference cannot properly track the location of these massive star-forming substructures past time of observation.

\begin{figure*}
    \centering
    \includegraphics[page = 1, width=0.95\linewidth]{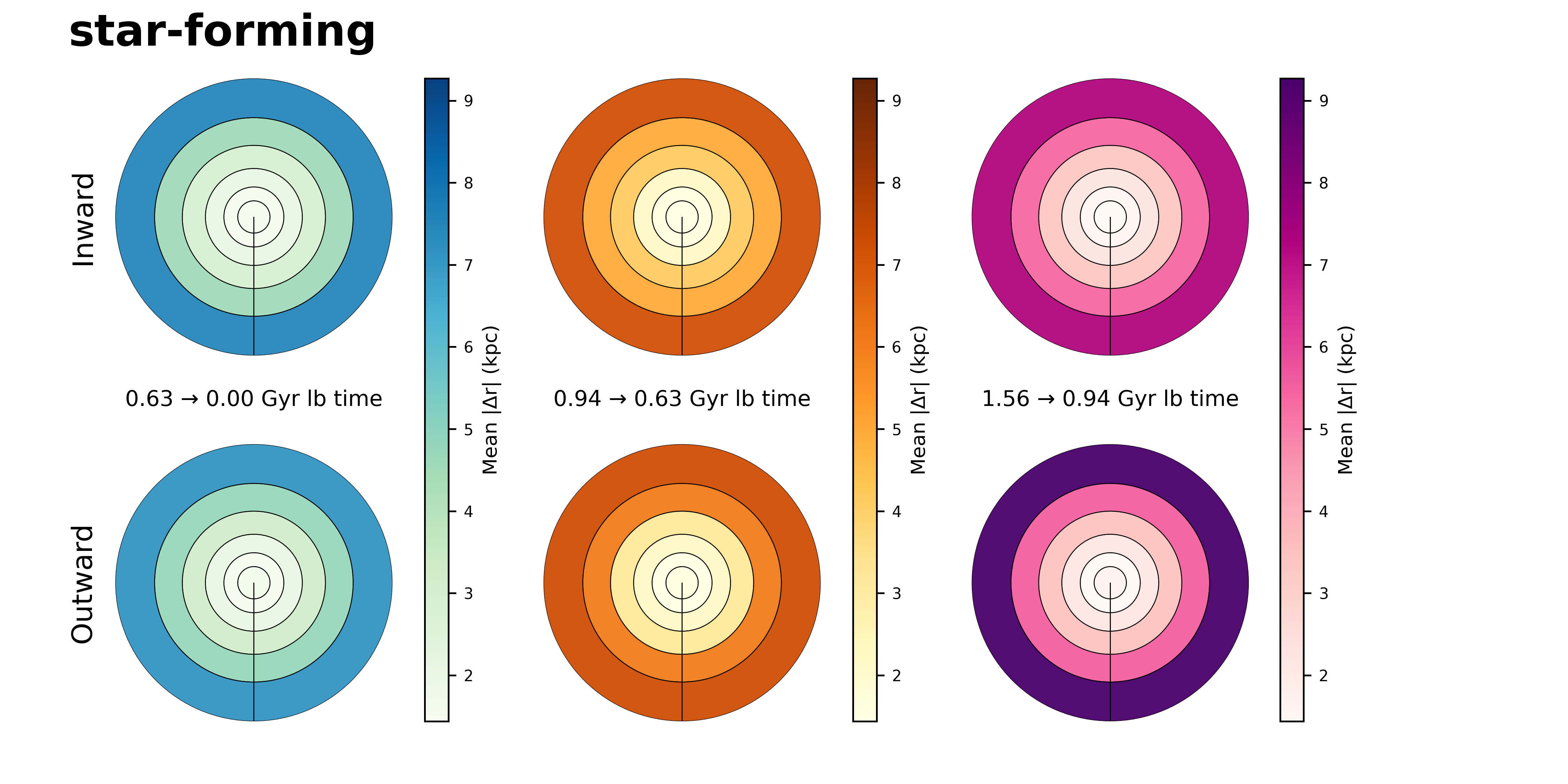}
    \includegraphics[page = 2, width=0.95\linewidth]{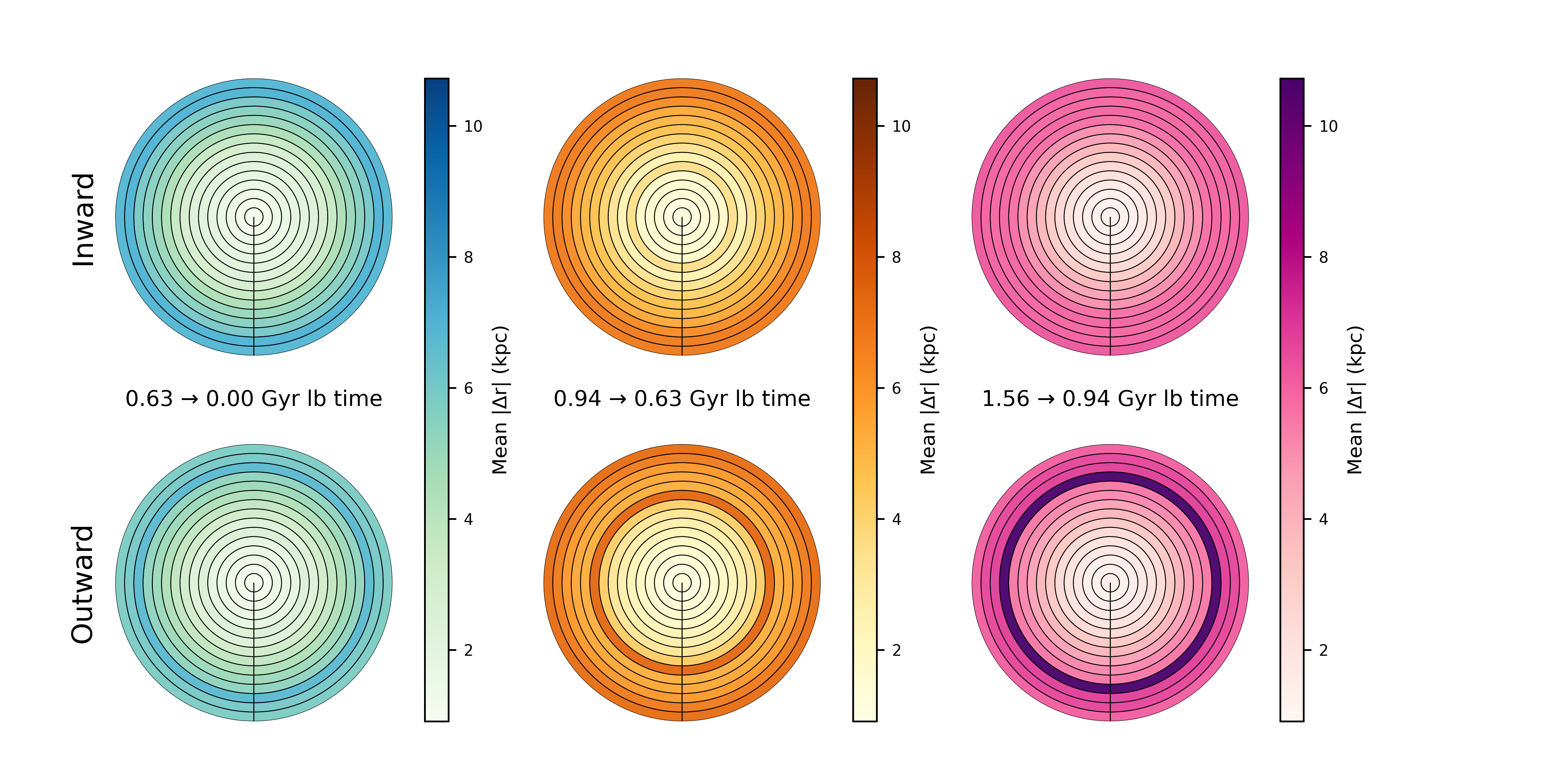}
    \caption{Mean radial distance travelled by bin-crossing star particles in the star-forming window, for variable-width bins (top) and fine 1 kpc bins (bottom), at each $\sim0.5$ Gyr step (columns). Rows show particles that moved into (Inward) or out of (Outward) each bin. In both schemes the distance grows with radius, from $\sim1.5$ kpc near the center to $\sim7$--9 kpc in the outer disc. The fine bins also pick out individual outer radii where particles jump much farther, up to $\sim11$ kpc in a single step, which we attribute to resonances; the wide outer bin averages these jumps out.}
    \label{fig:window3_migration_bin_dist_compare}
\end{figure*}

\begin{figure*}
    \centering
    \includegraphics[width=\linewidth]{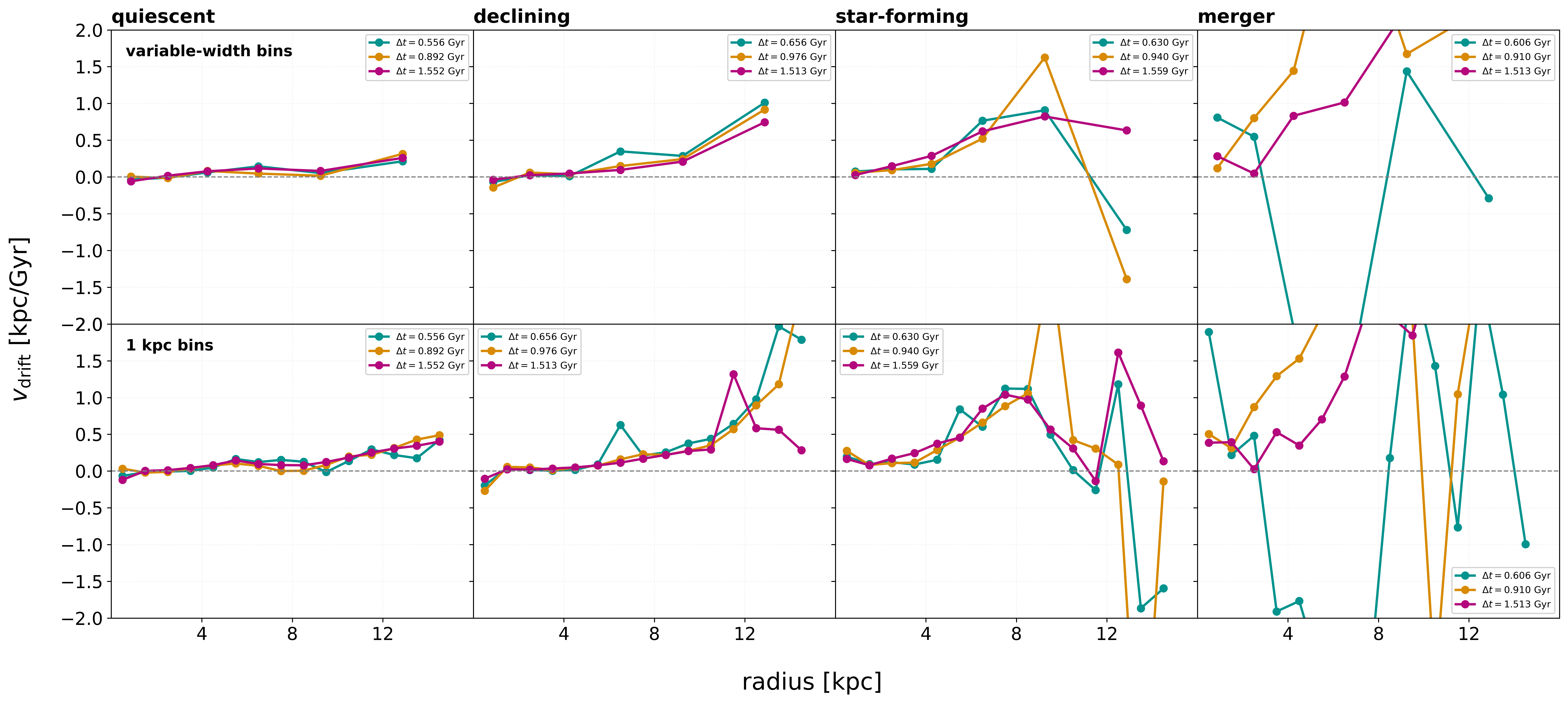}
    \caption{Bin-wise drift velocity $v_{\rm drift}$ (Eq.~\ref{eqn:drift}) as a function of galactocentric radius for each window, with variable-width bins (top) and fine 1 kpc bins (bottom). $v_{\rm drift}$ combines the fraction of a bin's particles that enter or leave it with the mean distance they travel, so positive values mean a net outward flow of stars. In every window the drift grows with radius, and variable-width bins reduce it at all radii compared with fine bins. In the star-forming window the drift changes sign beyond $r\sim9$ kpc, a signature of co-rotation, and in the merger window it is large at all radii.}
    \label{fig:drift}
\end{figure*}

To properly compare the radial advection per radial bin we define the bin-wise drift as such: 
\begin{equation}
    v_{\mathrm{drift}}(R_i,t)=\frac{f_\mathrm{out}d_\mathrm{out}-f_\mathrm{in}d_\mathrm{in}}{\Delta t}
    \label{eqn:drift}
\end{equation}
 where $f_{out} = \frac{N_{out}}{N_{bin}}$ and $f_{in} = \frac{N_{in}}{N_{bin}}$ which is the ratio of the number of star particles to have entered or exited the bin (out or in) divided by the total number of star particles in that bin. $d_{in}$ and $d_{out}$ are the mean distance traveled by star particles that have entered and exited the bin, and $\Delta t$ is the lookback time in Gyr between the epochs being compared. In this way, we can quantitatively compare the bulk drift between bins over each epoch and each window. This gives us the drift as a function of radius, which we show in Figure~\ref{fig:drift} for the variable and fine binning. 

 Across all temporal windows and binning prescriptions, the bin-resolved drift exhibits a systematic increase with radius. However, the magnitude of the drift in the outer regions varies with the level of star formation activity characteristic of each evolutionary window. During the quiescent phase, the outermost bins exhibit minimal drift, whereas, excluding the highly bin instable merger interval, the star-forming phase is characterized by maximal drift. Moreover, across all evolutionary windows, the variable-width binning scheme systematically reduces the influence of $v_\mathrm{drift}$ across bins relative to the fine-binning approach. It is worth noting here that the negative or positive values of $v_\mathrm{drift}$ show the bulk direction of stellar migration from bin to bin. During the quiescent and declining phases, stellar populations exhibit outward radial migration to larger galacto-centric radii. 
By contrast, in the star-forming phase, this trend is partially reversed: while star particles migrate outward up to $r\sim$9 kpc at a rate of approximately 1–1.5 kpc/Gyr during the first two epochs, at radii beyond this threshold the net motion reverses, with star particles migrating inward at comparable rates. This change in sign directly shows the resonance indicative of co-rotation as seen in \citep{Zhang2025}. This confirms that large near-resonance stellar clumps are the component of migration that SFH-based reconstructions struggle to mitigate as these are faster migrating and enclose a non-negligible fraction of stars. In such cases, this justifies limiting reconstructions to reduced lookback times ($\Delta t<1$ Gyr at $z<1$, i.e below the typical dynamical time of the disc) and accounting for kpc-level error on the position of sSFR peaks.




\section{Discussion}
\label{sec:discussion}

\subsection{The need for an adaptive radial binning scheme}

Throughout our study, we systematically compared two spatial binning schemes: (i) a fixed-radius binning with fine sampling, adopting a bin width of 1 kpc to match the typical spatial resolution of low-redshift IFU surveys; and (ii) an adaptive binning defined by spaxel sizes produced by Voronoi-like tessellation algorithms, as commonly implemented in spatially resolved SED fitting, with larger typical voxel size at larger radii.

In the majority of cases, we obtained improved performance by binning the data using the custom variable-width radial binning scheme (ii). It should be noted that bin widths were not initially chosen to account for stellar migration but were anchored to the mean spaxel size reported by \citet{Olsen2026}, in which the spaxel size was determined by the signal-to-noise ratio (SNR), $\chi^2$, and the point-spread function (PSF). However, it appears that these binning regions serendipitously scaled to account for the degree of dynamical mixing within the galaxy: the lower-density outer regions exhibit reduced SNR and therefore require larger spaxels, while also corresponding to the largest characteristic migration distances for star particles moving into or out of the outermost radial bin. Figure~\ref{fig:voxel_v_lit} shows the same median voxel radial extent versus radius as Figure~\ref{fig:voxelsize} from UVCANDELS, but now includes the average radial stellar drift at the third epoch ($\Delta t \sim 1.6$ Gyr) compared with literature values at $\Delta t = 3$ Gyr (twice our maximum lookback time), as a function of galacto-centric distance for all evolutionary windows.

This confirms that the adaptive binning of \citet{Olsen2026} effectively trends with the amplitude of stellar migration at increasing radius, and ensures that the average stellar drift in any voxel remains below voxel size across The window of observation (except in the case of major mergers, when the drift average amplitude reaches the scale of the entire galaxy). Of note, wider bins can accommodate different types of stellar migration, not only cold orbit changes but also hot migration processes, like outer stars with potentially higher eccentricity, which are still more likely to remain in their initial wider bin. Note that for our purposes, it is irrelevant whether the star particles underwent blurring (eccentricity increase) or churning (orbit jumps), because the changes we observe in the radial profiles depend only on each star particle’s initial and final bin locations. 

In conclusion, while voxels from \cite{Olsen2026} serendipitously accounted for the bin-to-bin drift, they now point to the necessity of such a migration-aware binning scheme in all SFH-based reconstructions. 

\subsection{ h277 as an upper envelop of radius-dependent migration strength}

\begin{figure}
    \centering
    \includegraphics[width=0.95\linewidth]{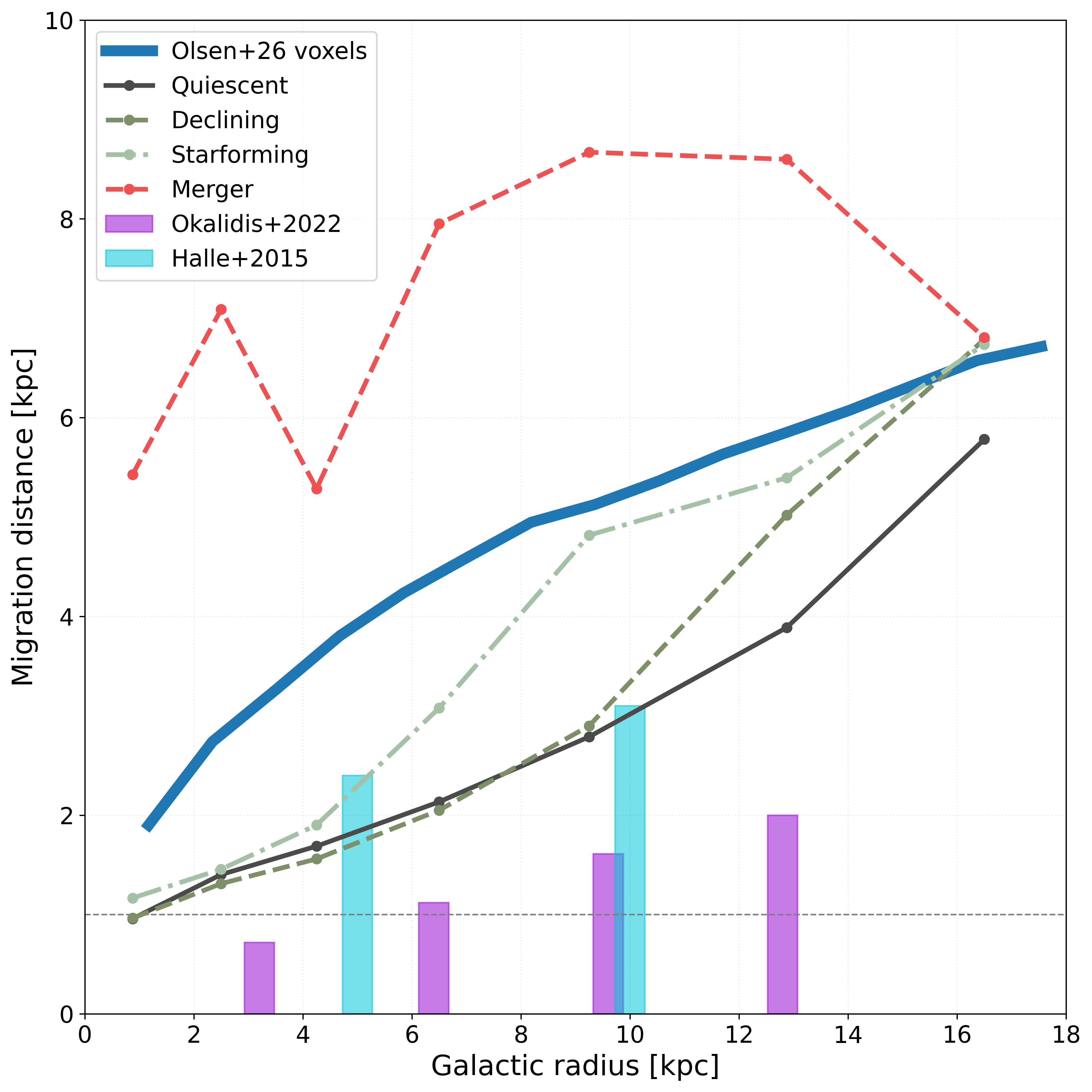}
    \caption{Radial migration distance as a function of galactocentric radius, compared with the voxel size. Dashed and dotted lines show the mean distance travelled by h277 star particles that changed bin in each window at the epoch corresponding to $\Delta t \sim 1.6$ Gyr. Cyan and purple bars show the migration over 3 Gyr reported by \citet{Halle2015} and \citet{Okalidis2022}. The solid blue line is the median voxel size from Fig.~\ref{fig:voxelsize}, and the dotted horizontal line marks the 1 kpc width of the fine bins. Outside the merger window, migration stays below the voxel size at every radius but exceeds the width of the fine bins, so variable-width bins contain most migration while fine bins do not.} 
    \label{fig:voxel_v_lit}
\end{figure}

Theoretical predictions from \cite{Okalidis2022} and \cite{Halle2015} are also added to Figure~\ref{fig:voxel_v_lit} as colored bars. Note that, while they report the rate of radial migration as a function of galactic radius, both report mixing over $\Delta t=3$ Gyr as compared to $\Delta t \sim 1.5$ Gyr for this work. But the lack of bar and lack of mergers respectively suggest the mixing would be less than for \textbf{h277}. The \cite{Halle2015} numbers are derived from an idealized, isolated N-body/SPH model of a single barred, Milky Way–like disk rather than a full cosmological setup. Their configuration is intended to separate churning from blurring by evolving and recording both the instantaneous radius and the guiding radius. In this framework, migration is expressed as the stellar mass fraction that moves beyond a specified $|\Delta R|$ threshold, evaluated in radial zones associated with the bar’s resonances. Because the disk evolves in isolation, the inferred migration reflects purely secular, bar-induced redistribution, excluding cosmological accretion, mergers, and any response of the dark-matter halo. \cite{Okalidis2022} report results from the Auriga cosmological magneto-hydrodynamical zoom-in simulations. The cited $\sigma_{migr}$ values are selected from a weakly barred system, and are defined as the 16th–84th percentile range of each star’s change in galacto-centric radius between birth and the final snapshot. Since this halo does not host a prominent bar, these migration estimates represent a lower bound compared with \textbf{h277} or more merger-perturbed Auriga galaxies. Our estimated migration rate is consistently higher than those reported by \cite{Halle2015} and \cite{Okalidis2022} across all windows.

Overall these results support the idea that the strongly barred (and clumpy/perturbed at earlier times) \textbf{h277} disc represents an excellent case to test the upper bounds of stellar migration (amount, span) in  disc galaxies and is therefore particularly adapted to the development of a conservative radial binning scheme for spatially resolved SFH. Indeed, \textbf{h277} holds the advantage of being a disc galaxy that displays contrasted, clumpy, transient  spiral arms across our test windows, ensuring a high degree of churning at the co-rotation and Lindblad resonances \citep[changes the angular momentum of star particles without heating their orbits: ][]{SellwoodBinney2002, Roskar2008}. Since multiple transient patterns with different pattern speeds coexist in the disk, their co-rotation radii, hence the resulting migration, span a large fraction of the disk \citep{Roskar2012}, and spiral arms co-rotating with the disk can drive efficient migration at all radii \citep{Grand2012}. 

In addition, bars contribute additional migration near their own co-rotation and through the overlap of bar and spiral resonances \citep{MinchevFamaey2010, Minchev2011}, with an efficiency that depends on both the strength of the bar and the dynamical temperature of the disk \citep{Brunetti2011} but typically remains below that of spiral arms, especially at larger radii. In that regard, \textbf{h277} displays episodes of relatively rapid bar growth/slowing (consecutive to inflows) then bar decay/speed up across our non-merging windows of observation, therefore maximizing the amount and amplitude of bar-driven migration with radius as the co-rotation radius sweeps across the disc \citep{Zhang25}.

Meanwhile, external perturbations, like inflows or satellite passages, preferentially affect the outskirts of the galaxy \citep{Quillen2009}. As a result,  migration is boosted at large radii in higher redshift disks still perturbed by their accretion history \citep{Bird2012}. Similarly, for a given level of velocity dispersion, blurring produces larger radial excursions at larger radii \citep{BinneyTremaine2008}. All these results therefore support the relative generality of the increasing trend for migration amplitude with radius seen in \textbf{h277}. Finally, the ability of \textbf{h277} to form star particles from dynamically cold gas \citep{Bird2021} is especially relevant in this context, as migrating stars are preferentially drawn from the kinematically coldest populations \citep{VeraCiro2014}. Simulations in which young stars are born too hot would therefore underestimate churning, which is precisely the component of migration most likely to bias spatially resolved SFHs. 

Therefore, we conclude that the voxels size increase with radius derived from \citet{Olsen2026} constitutes an excellent upper envelope for the mean distance traveled by migrating stars in discs as a function of radius (across non-merging cases), as can be seen in Figure~\ref{fig:voxel_v_lit}. Irrespective of S/N constraints, we therefore recommend this scaling for future SFH-derived profile reconstructions.

These values are also in line with existing prediction of the amplitude of migration expected over the $\Delta t \lesssim 1.5$ Gyr lookback time. For instance, if migration is diffusive, its spread should grow as $\sqrt{t}$. The age-metallicity distribution of the low-$\alpha$ disk of the Milky Way implies $\sigma_{RM}\simeq 3.6$ kpc over 8 Gyr \citep{Frankel2018}, which corresponds to $\sim 0.9$, $1.3$ and $1.6$ kpc after $\Delta t = 0.5$, $1$ and $1.5$ Gyr respectively, and this redistribution is dominated by changes in angular momentum (churning) rather than radial heating (blurring) \citep{Frankel2020}. In the FIRE-2 simulations of Milky-Way mass galaxies, the spread in radial displacement $\sigma(\Delta R)$ increases with stellar age only up to $\sim 3$ Gyr, and saturates at $\sim 2$ kpc for older stars \citep{Bellardini2026}. In the Auriga simulations, the strength of migration varies between 1 and 4 kpc as a function of radius and stellar age. Migration in \textbf{h277} is of the same order: in the quiescent window, star particles that cross a bin edge travel on average $\sim 1.4$ kpc in the central bin, i.e. less than its 1.75 kpc width, and up to $\sim 5$ kpc in the outermost bin, where fewer than $4\%$ of star particles cross over each $\sim 0.5$ Gyr step. However, these mean distances are computed for bin-crossing star particles only, hence are biased high compared to the population-wide $\sigma(\Delta R)$ reported in the literature. The drift-based estimate presented in Figure~\ref{fig:voxel_v_lit} therefore provides a more direct comparison.

\subsection{Limitations in the perturbed cases}

Strongly perturbed cases (by inflows, clumps, mergers) present interesting departure from the more settled cases of window 1 and 2. Indeed, while our results for the clumpy, gas-rich, star forming window show that the mean traveled distance still increases with radius as in more settled phases, the amplitude of this migration for bin-crossing star particles can now reach $\sim 7-9$ kpc in the outer disc. This in line with strong co-rotation scattering near the prominent, clumpy spiral structures \citep[see Fig.~\ref{fig:window3_migration_bin_dist_compare} but also:][]{Roskar2012, Grand2012} also visible from the change in sign of $v_\mathrm{drift}$ beyond $r\sim 9$ kpc on Fig.~\ref{fig:drift}. Our suggested radial binning scheme still allows to mitigate most of the effects of the outer disc on median sSFR radial profiles out to $\Delta t\sim 1$ Gyr but leads to large errors at larger lookback times. Beyond the outer disc, the positions of large star-forming clumps can also be offset by $1-2$ kpc by 1 Gyr lookback times, and the impossibility to time the initial emergence of such clumps offers further incentive to restrict lookback times below typical survival time of clumps, which decreases with redshift \citep{Mandelker16}.

What's more, as many similar discs, \textbf{h277} grows inside-out \citep{Bird2021}: the renders in Figure~\ref{fig:globalsfh}, all shown at the same scale, display a more compact stellar disk in the star-forming window than in the declining and quiescent windows, indicating a growing disk. Resolved SFHs attribute past star formation to the spatial extent of the galaxy at the time of observation, without accounting for later loss of compactness. This contribute to shifting inferred star-forming features to larger radii, consistent with the 1-2 kpc outward offset of the inferred scatter peaks (Appendix~\ref{apendix:offset}), and both increase with lookback time. This further justifies limiting reconstructions to $\Delta t \lesssim 1$ Gyr for clumpy, highly star-forming galaxies.

\begin{figure}
    \centering
    \includegraphics[width=0.95\linewidth]{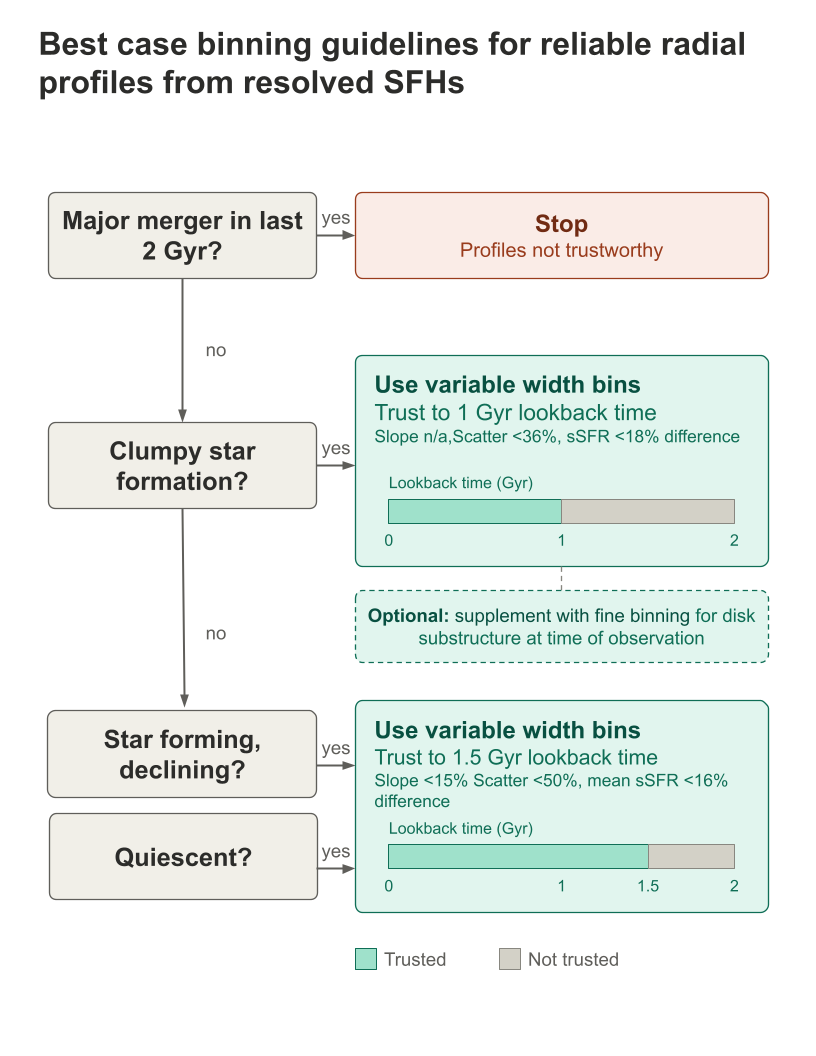}
    \caption{Recommended steps for binning photometric data and choosing how far back to trust radial profiles from resolved SFHs, based on the average percent difference in slope, scatter and median-sSFR biases for each window in Figs.~\ref{fig:slopescatter4windows} and \ref{fig:scatterperbin}. Variable-width bins that grow with radius give the most reliable profiles. With them, profiles can be trusted to $\sim1.5$ Gyr lookback time for quiescent or declining galaxies and to $\sim1$ Gyr for actively star-forming ones. Finer bins are still useful for studying disc substructure at the time of observation. We recommend against reconstructing profiles for galaxies with a major merger in the past 2 Gyr.}
    \label{fig:flowchart}
\end{figure}

Major mergers, however, fall outside of this secular regime altogether. In cosmological disks, mergers at high redshift are the dominant driver of large-scale migration before the bar takes over at later times \citep{Minchev2013}. This is consistent with the bin-crossing fractions reaching $\sim 60\%$ and displacements reaching $\sim 13$ kpc in our merger window. No choice of radial binning can absorb migration on the scale of the entire galaxy, which is why we recommend against SFH-based reconstruction of radial profiles for galaxies displaying signs of a recent major merger, such as strong tidal features and asymmetries \citep{Wen2016,Desmons2026} or, if IFU data are available counter-rotating cores and higher-order kinematic moments \citep{VdS17}. Our recommendations are summarized for convenience in Figure~\ref{fig:flowchart}.


\section{Conclusions}
\label{sec:conclusions}
We used the \textbf{h277} MW mass simulation to test the accuracy of past radial profiles of sSFR reconstructed from spatially resolved star formation histories. We test in particular the ability of such reconstructions to recover the slope of profiles -- hence discriminate between inside-out vs. outside-in evolution -- out to 1.5 Gyr lookback time, with different radial binning schemes: regular, 1 kpc wide bins (as available in large IFS surveys) and radial bins with width increasing with radius, matched to average spaxel size from \cite{Olsen2026}  at each radius  (SED derived SFHs from multi-band photometry). This constitutes a much needed validation for future large photometric surveys like LSST/Rubin. We reconstruct SFHs in several evolutionary windows: a quiescent phase, a declining  but still star-forming phase, an active, clumpy star-forming phase, and a major merger. For each window we generated ``truth'' sSFR radial profiles at lookback times of $\Delta t\sim$ 0.5, 1, and 1.5 Gyr, and compared it to the sSFR profile derived from SFHs on the reference snapshot. Our main findings are:

\begin{itemize}
    \item \textbf{Variable-width bins can scale with radius and should be made adaptive to migration rates.} Spaxel sizes set by SNR, $\chi^2$, and the PSF grow with galacto-centric radius, and so does the typical distance stars migrate. As an example, our specific binning from \cite{Olsen2026} proves to absorb most migration effects across different levels of star formation. Indeed, variable-width bins minimize the net bin-wise drift, $v_{\rm drift}$, at all radii relative to fixed 1\,kpc bins. Migration in the quiescent, declining star forming, and clumpy star-forming windows stays below the median spaxel size at every radius. This is not the case with regular 1-kpc bins. 
    
    \item \textbf{Inner-disk mixing is mostly local.} At small radii, apparent mixing is dominated by short-range crossings between adjacent bins, which largely cancel out. The largest displacements occur in the outer disk, where the widest variable-width bin spans the region of strongest mixing and keeps blurred stars in their original bin.

    \item \textbf{Variable-width binning therefore recovers radial profiles more accurately.} Both schemes recover the radial sSFR slope with $<20\%$ error but the variable-width scheme displays $<12\%$ across all cases. Over $\Delta t\sim 1$ Gyr of lookback time, however, fixed 1 kpc bins systematically produce higher errors on the mean binwise sSFR compared to variable-width bins that recover it with $<20\%$ error. Fine bins also under-perform at reproducing the scatter in sSFR radial bins  compared to variable width binning, although both show a tendency to underestimate the scatter by $>30\%$.

    \item \textbf{Radial location of spiral-associated large clumps and star forming features is preserved at $z<1$.} While the strength of these features can be over or underestimated due to migration between bins, we recover the location of these features down to 1-2 kpc using their current peak in sSFR, for both binning schemes, as long as $\Delta t <1$ Gyr in our analysis. This is owing to their relatively low-migration rate, compared for instance to galaxies above $z>2$ undergoing stronger accretion and faster clump migration. 

    \item \textbf{The presence of highly star forming features impacts recovery of past epochs.} For quiescent and declining (but still star forming) galaxies, variable-width binning keeps errors below 16\% out to $\sim$1.5\,Gyr lookback time, owing to their relatively smooth, monotonous radial profiles. For actively star-forming galaxies with clumpy features, profiles are reliable only to $\sim$1\,Gyr; at 1.5\,Gyr the sSFR error reaches 48\% under either scheme due to clump migration, and the scatter is most strongly underestimated.

    \item \textbf{Recent major mergers cannot be corrected by radial binning.} During a major merger, stars spread across all radial bins, and no binning scheme recovers the true profile. We recommend against building radial profiles from resolved SFHs for galaxies with evidence of a major merger in the past 2\,Gyr.

    \item \textbf{Recommendation.} For galaxies without a recent major merger, we recommend radial bins with width increasing with galactocentric radius so as to scale with migration, as shown in Figure~\ref{fig:voxel_v_lit}. Profiles should be interpreted out to $\sim$1\,Gyr lookback time for clumpy, star-forming systems and $\sim$1.5\,Gyr for quiescent or relatively smooth systems (Figure~\ref{fig:flowchart}).
\end{itemize}

With upcoming large surveys, vast amounts of photometric data will soon be available. New observations of galaxies in large statistical samples promise to reveal a deeper understanding of how and where galaxies shut off star formation. Comparing age-regressed radial profiles from resolved SFHs will be a reliable and illuminating method for exploring the frequency and timescales of inside-out vs outside-in growth and quenching.  

\section{Acknowledgments}
\label{sec:acknowledgements}

The core scientific team is comprised of Charlotte Olsen, Charlotte Welker and Ena Chia. CO led the analysis and produced most figures. EC implemented an early version of the pipeline and produced some figures. CW designed the study and methodology and supervised the team. CO and CW wrote the initial draft and implemented comments.  All authors commented on the draft.
CO and CW acknowledge the support of the LSST-DA Catalyst Fellowship funded by the Templeton Foundation. CW acknowledges the support of the NSF LEAPS-MPS Award AST-2316862 and the LSST-DA Early Science award funded by the Simons foundation. EC acknowledges the support of the AstroCom NYC fellowship funded by NSF awards AST-2219090 and AST-1831412 and of the Emerging Scholars program at City Tech. The authors would like to thank Alyson Brooks, and Kartheik Iyer for their very helpful feedback. No generative AI tool was used in the design, data preparation, figure preparation, interpretation or writing of this study.  An AI coding copilot was used for low-assistance coding tasks during the analysis: code cleaning, syntax checking, minor re-formatting, commenting.

\bibliographystyle{aasjournalv7}
\bibliography{author}

\appendix
\vspace{1cm}
\section{Fine binning in radial profiles}
\label{appendix:finerp}

\begin{figure*}
    \centering
    \includegraphics[width=0.95\linewidth]{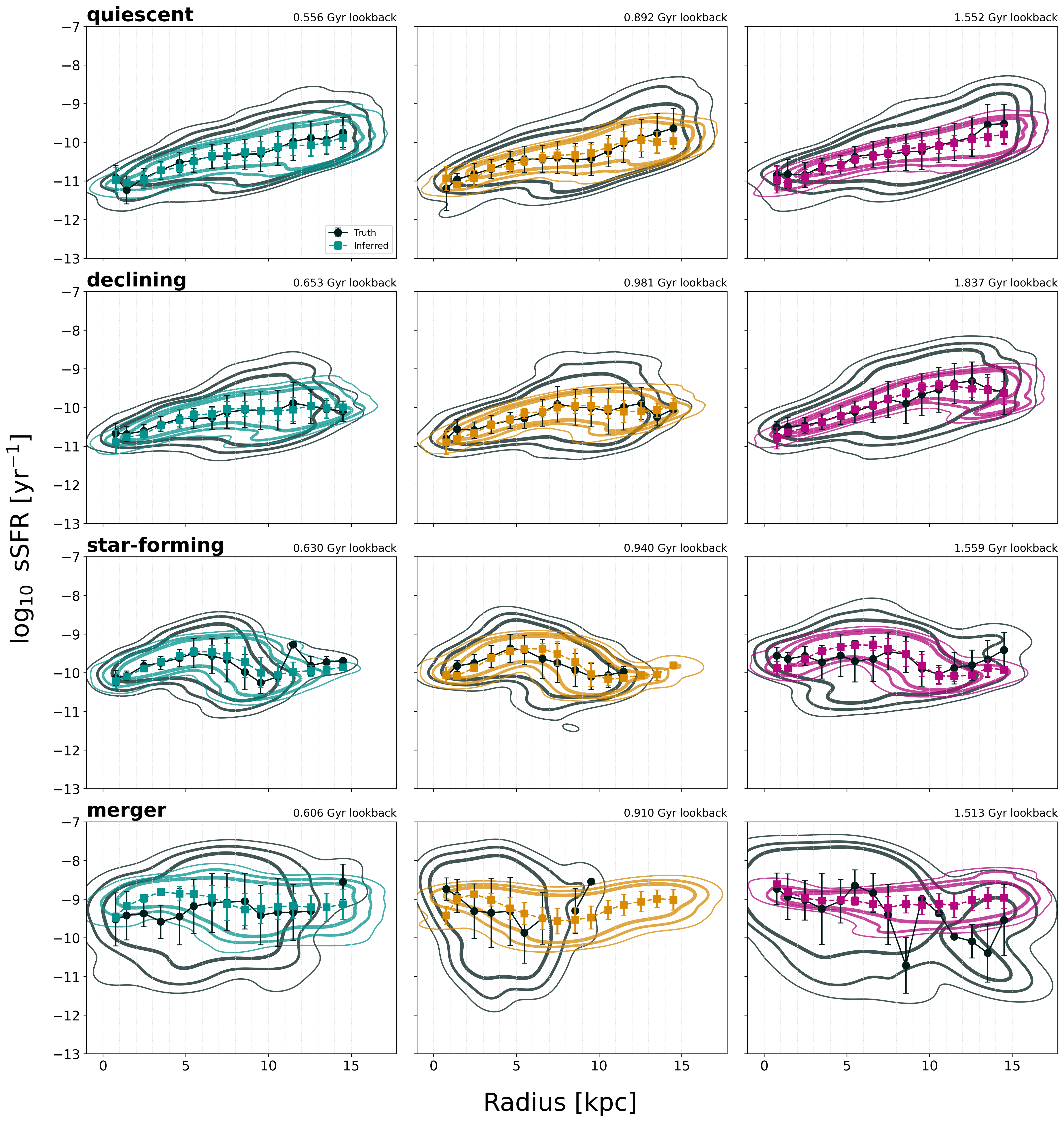}
    \caption{As Fig.~\ref{fig:radialprofiles}, but with fine 1 kpc bins, similar to the spaxel size of IFU surveys. Outside the merger window, the inferred median profile still follows the truth out to $\sim1$ Gyr lookback time. However, the inferred profiles miss the sharp sSFR peak from a star-forming clump at $r\approx12$ kpc in the star-forming window, and they diverge from the truth in the outskirts from $\Delta t\approx1$ Gyr. By $\Delta t\approx1.5$ Gyr this divergence reaches in to $r\approx5$ kpc in the declining and star-forming windows.}
    \label{fig:fineradialprofiles}
\end{figure*}

Fig.~\ref{fig:fineradialprofiles} displays sSFR radial profiles obtained with the same dual methodology as in Figure~\ref{fig:radialprofiles} but using a constant bin width of 1 kpc, akin to typical binning in large IFU surveys such as SAMI\citep{Croom12} or Hector\citep{Bryant2024}. Qualitatively, the reconstruction with this binning also recovers radial profiles overall, across most evolutionary stages but the major merger, at least up to 1 Gyr lookback time.

However, at the bin level, discrepancies between the inferred and ground-truth profiles are already detectable on these median sSFR profiles. This is most notable in the star-forming window, where a clump at $r\sim12$kpc generates a sharp increase in sSFR  at $\Delta t \sim 0.5$ Gyr lookback time, that is completely missed by the inferred profile. Slope Biases between the actual and inferred median radial profiles also appear in the outskirts of the galaxy, causing inferred and ground-truth trends to diverge significantly as early as the $\Delta t\sim 1$ Gyr epoch.

At $\Delta t\sim 1.5$ Gyr, visible biases start to reach the inner parts (down to 5 kpc) of the galaxy for both the declining and star-forming windows. Comparing the narrowing of the scatter in the inferred profiles can not be done easily by eye. But Figure~\ref{fig:slopescatter4windows} shows a direct comparison between the underestimation of both methods, and with exception of the first epoch of the star-forming window which matches that of the variable width bins, the fine binning is systematically underestimating the spread of sSFR for all other epochs and windows.  This shows that, irrespective of instrumental precision, fine binning may hinder the accuracy of sSFR profile reconstruction compared to adaptive radial binning.

\section{Peak offset in the star-forming window}
\label{apendix:offset}

\begin{figure*}
    \centering
    \includegraphics[width=0.95\linewidth]{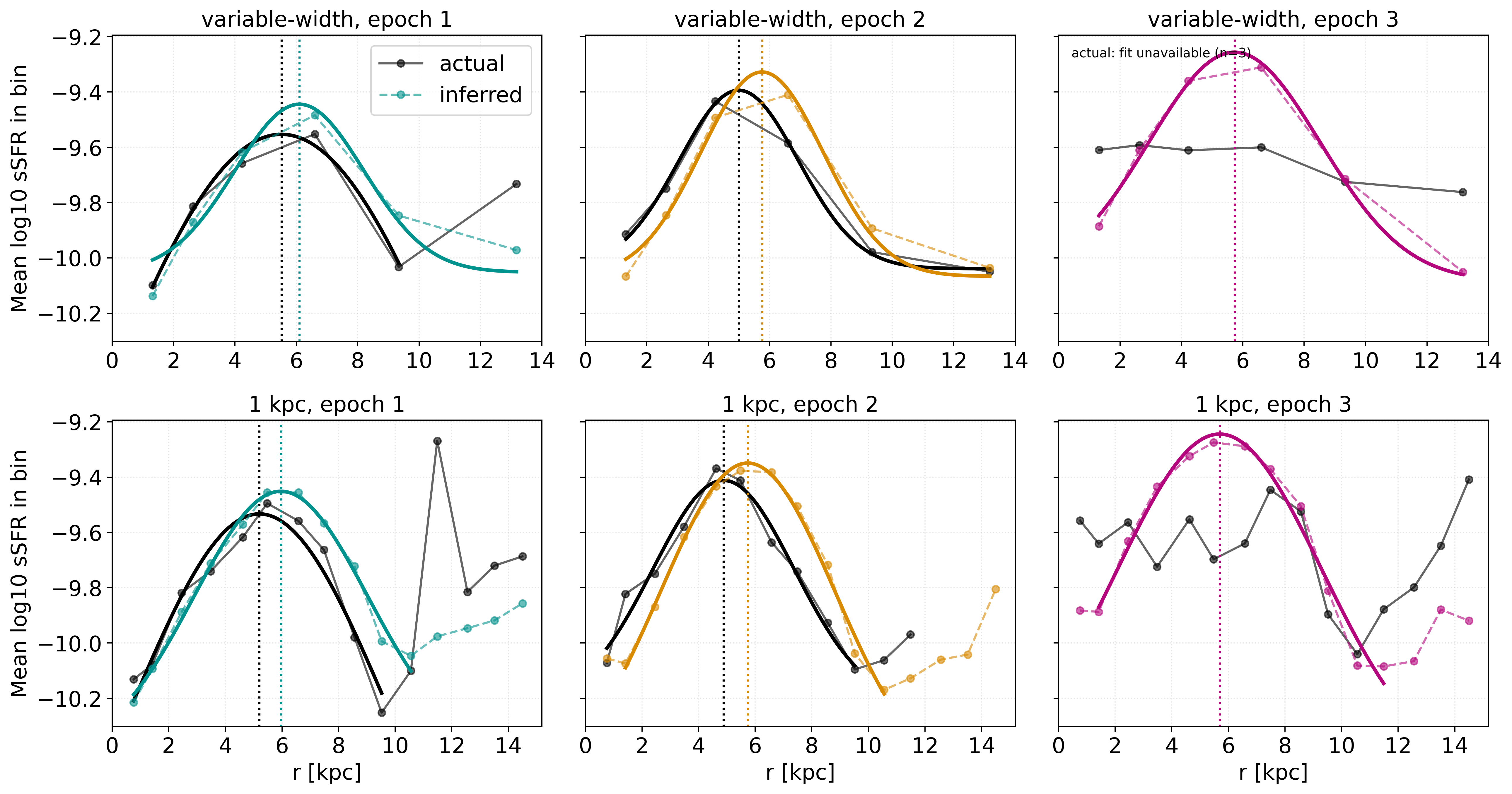}
    \caption{Radial sSFR profiles in the star-forming window at each epoch (columns), for variable-width bins (top) and fine 1 kpc bins (bottom). [Solid] lines show the true profile and [dashed] lines the inferred profile. Bold curves are Gaussian fits to each profile's peak, and vertical dashed lines mark the fitted peak radii. In the first two epochs both binning schemes recover the radius of the sSFR peak, caused by star-forming clumps in the spiral arms, to within 1 kpc. At $\Delta t\approx1.5$ Gyr the true profile has become flat, but the inferred profile still shows a peak.}
    \label{fig:peakoffset}
\end{figure*}

\begin{figure*}
    \centering
    \includegraphics[width=0.95\linewidth]{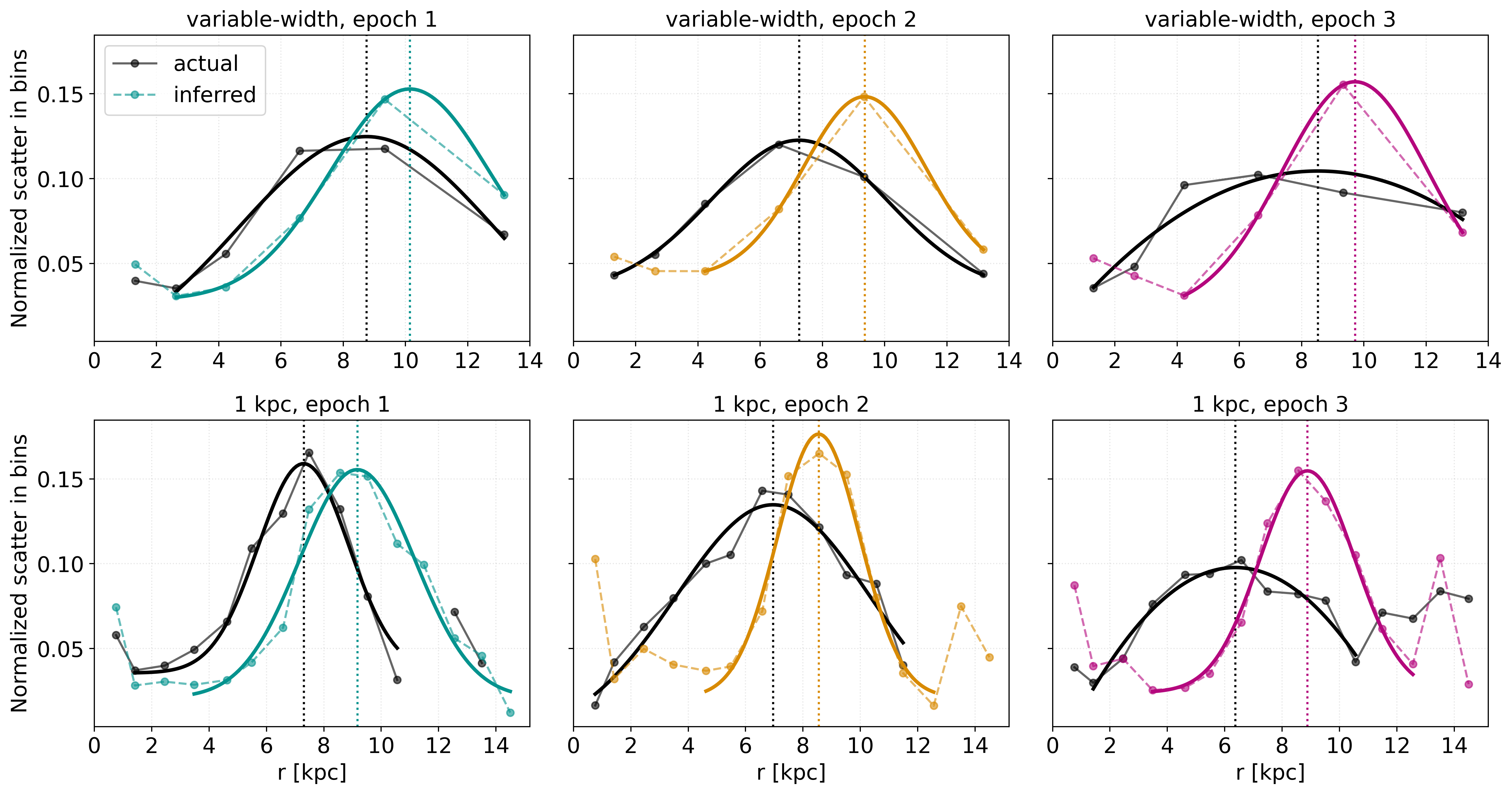}
    \caption{Normalized scatter $\sigma_{norm}$ vs. galactocentric radius in the star-forming window at each epoch (columns), for variable-width (top) and fine 1 kpc (bottom) bins.  Color indicates the inferred distribution with black as actual. Bold lines are the gaussian fits, and the vertical dashed lines show the fitted peaks. The inferred scatter peak is offset by 1--2.5 kpc (mean 1.4 kpc variable, 2.1 kpc fine).}
    \label{fig:peakoffset_scatter}
\end{figure*}

Figure~\ref{fig:peakoffset} and~\ref{fig:peakoffset_scatter} display the mean sSFR per bin and normalized scatter of the sSFR respectively (both actual and inferred) as a function of radial distance from the center of the galaxy for the star-forming window, which was shown to display non-monotonous sSFR profiles. Bold solid lines indicate gaussian fits to bin-wise values. The top row shows results for variable (SED-like) binning while the bottom row shows results for 1-kpc wide regular radial binning (IFU-like).

In Figure~\ref{fig:peakoffset} we fit a gaussian to the radial profiles in the star-forming window for the purpose of determining the offset of the peak in sSFR seen in the actual and inferred radial profile. We see that the offset in the peak is $<1$kpc and $\sim 1$ kpc in the first and second epoch respectively while the inferred radial profiles for epoch 3 still assume a peaked radial profile where the actual radial profile is flat. This shows that, at our redshift of observation ($z<1$) large spiral-associated clumps can be located in the outer spiral arms using to within 1 kpc of their actual past radial location up to 1 Gyr lookback time, using current peaks in sSFR proxy. This is because they migrate on longer timescales than high-z clumps from violent instabilities. The fine binning scheme offers no gain in resolution, and performs marginally worse in the first epoch compared to the variable width binning.

Qualitatively the overall, unimodal radial trend of the sSFR scatter is recovered across lookback times and binning schemes: increasing from the center of the galaxy out to $r\sim 6-9$ kpc and declining thereafter. However a systematic offset of$\sim 1-2$ kpc in the location of the scatter peak is also detected across all epochs.
On average, the offset is minimal for the variable bins (mean $\bar{\Delta r} = 1.4$ kpc) and maximal for the regular kpc-wide binning (mean $\bar{\Delta r} = 2.1$ kpc). The variable binning also maintains a better estimation of the trend's width at half-maximum across epochs (within 20$\%$ precision up to 1 Gyr), while the regular kpc-wide binning only performs well at $\Delta t$ 0.5 Gyr lookback time but quickly degrades at later times ($>50\%$ error) . Central and outer values are also better recovered with variable-width binning across all epochs.

Overall it confirms that, in the context of stellar migration, finer binning can perform worse than S/N driven adaptive binning typical of photometric reconstructions.

\section{Amplitude and prevalence of radial migration across evolutionary stages}
\label{appendix:migration}

\subsection{Across Variable bins}

In the following, we reproduce the analysis of Fig.~\ref{fig:window1_migration} but for the three other evolutionary windows:  decreasing star formation, star forming and merger. In each series of maps we quantify the fraction of stars migrated in and out of variable (photometric like) bins (top two rows), and the mean radial distance traveled by those bin-crossing stars across 0.5 Gyr timesteps.

\begin{figure*}
    \centering
    \includegraphics[page = 1, width=0.95\linewidth]{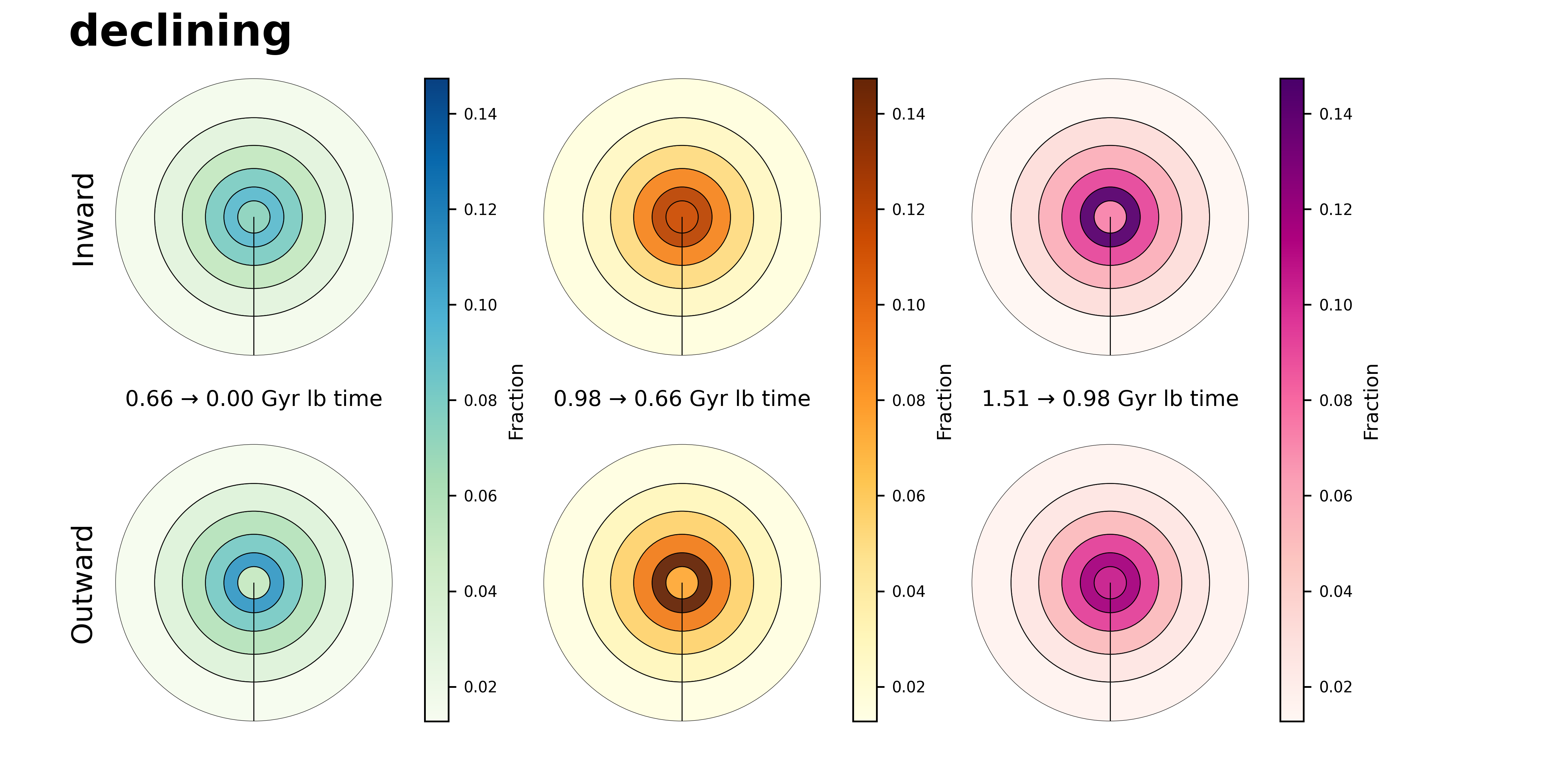}
    \includegraphics[page = 2, width=0.95\linewidth]{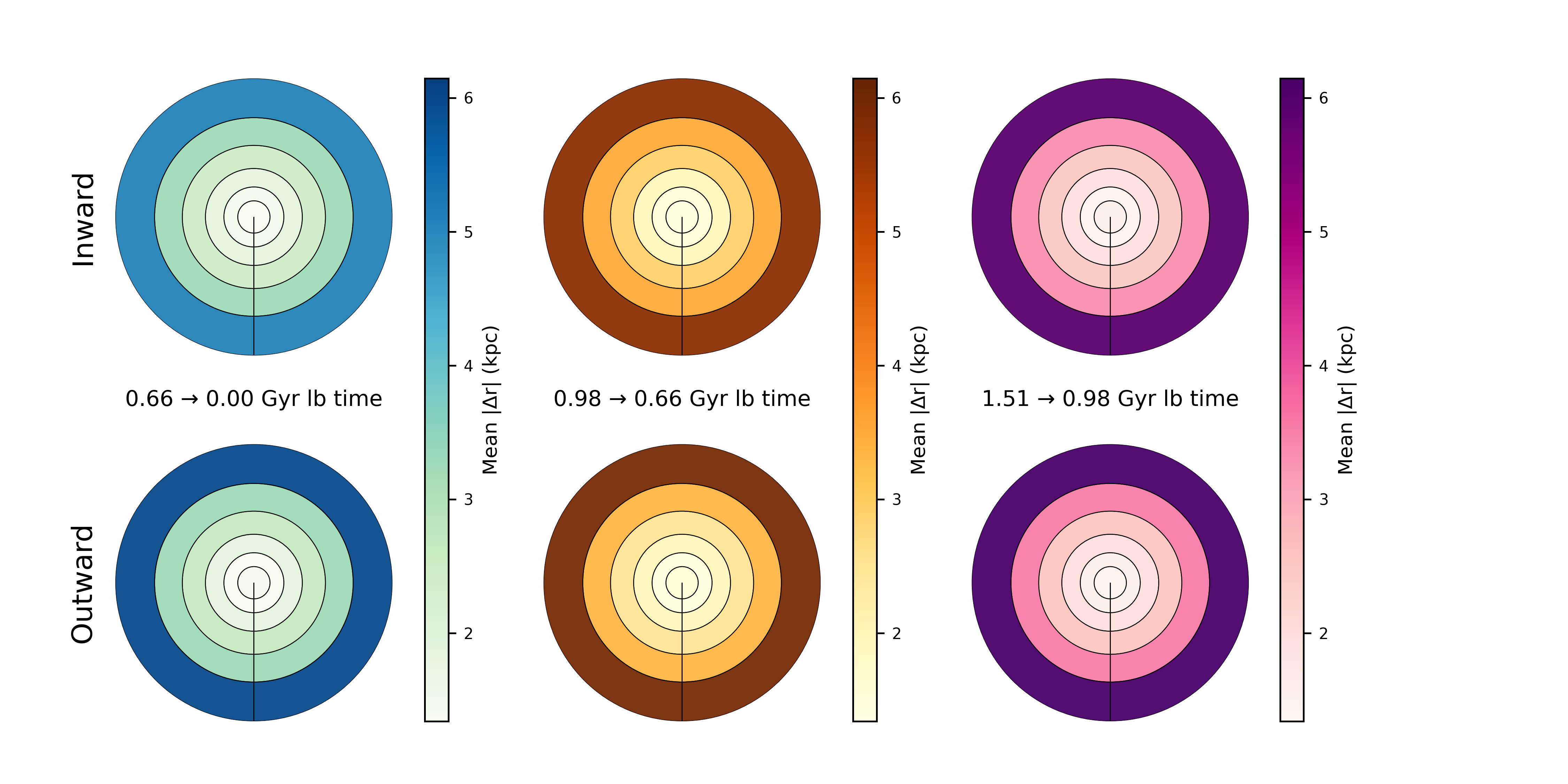}
    \caption{Window 2 (decreasing star formation): We see the same behavior as in the quiescent window (see Figure~\ref{fig:window1_migration}), where the fraction of stars moving into bins is equal to those moving out, with most bin crossing happening in the inner bins. The small number of stars that have moved into or out of the widest bin which is furthest from the galactic center are the ones that have on average traveled $\sim 6 kpc$, which is greater than the largest mean difference traveled in the quiescent window.}
    \label{fig:window2_migration}
\end{figure*}

Fig.~\ref{fig:window2_migration} focuses on the decreasing star formation window. It is apparent from the two top rows that the fraction of stars migrating across bins over $\Delta t=$ 0.5 Gyr remains extremely small ($<4\%$) in all bins but the one right around the center where it can reach $15\%$ at large lookback times. In this circum-central bin, the succession of inward migration dominance at $\Delta t= 1.5$ Gyr, followed by outward migration dominance at $\Delta t= 1$ Gyr is reminiscent of a cycle of inflow and AGN feedback outflow.

However, the bottom rows reveal that, even in this feedback dominated bin, the mean radial distance traveled by bin-crossing stars remains below the width of the central bin at 1.4 kpc. This is not surprising as gas accretion/outflows only indirectly affect the stellar radial distribution through minor modifications of the gravitational potential (gravitational steepening or puffing-up). Migration is therefore limited to minor "cold" squeezing or expansion of stellar orbits at the boundary between bins, unlikely to change the broader star formation profiles.

Overall, with the adaptive radial binning of \citep{Olsen2026} typical of photometric SED fitting reconstructions, the mean distance traveled by stars remains below or equal to their bin size across the face of the galaxy  (see Fig.~\ref{fig:voxel_v_lit}) and the mean distance traveled by bin-crossing stars ensure they do not cross more than one bin. In conclusion, variable binning ensures that fractions of crossing stars remain low and their drift is limited in span in our window of SFH reconstruction.

\begin{figure*}
    \centering
    \includegraphics[page = 1, width=0.95\linewidth]{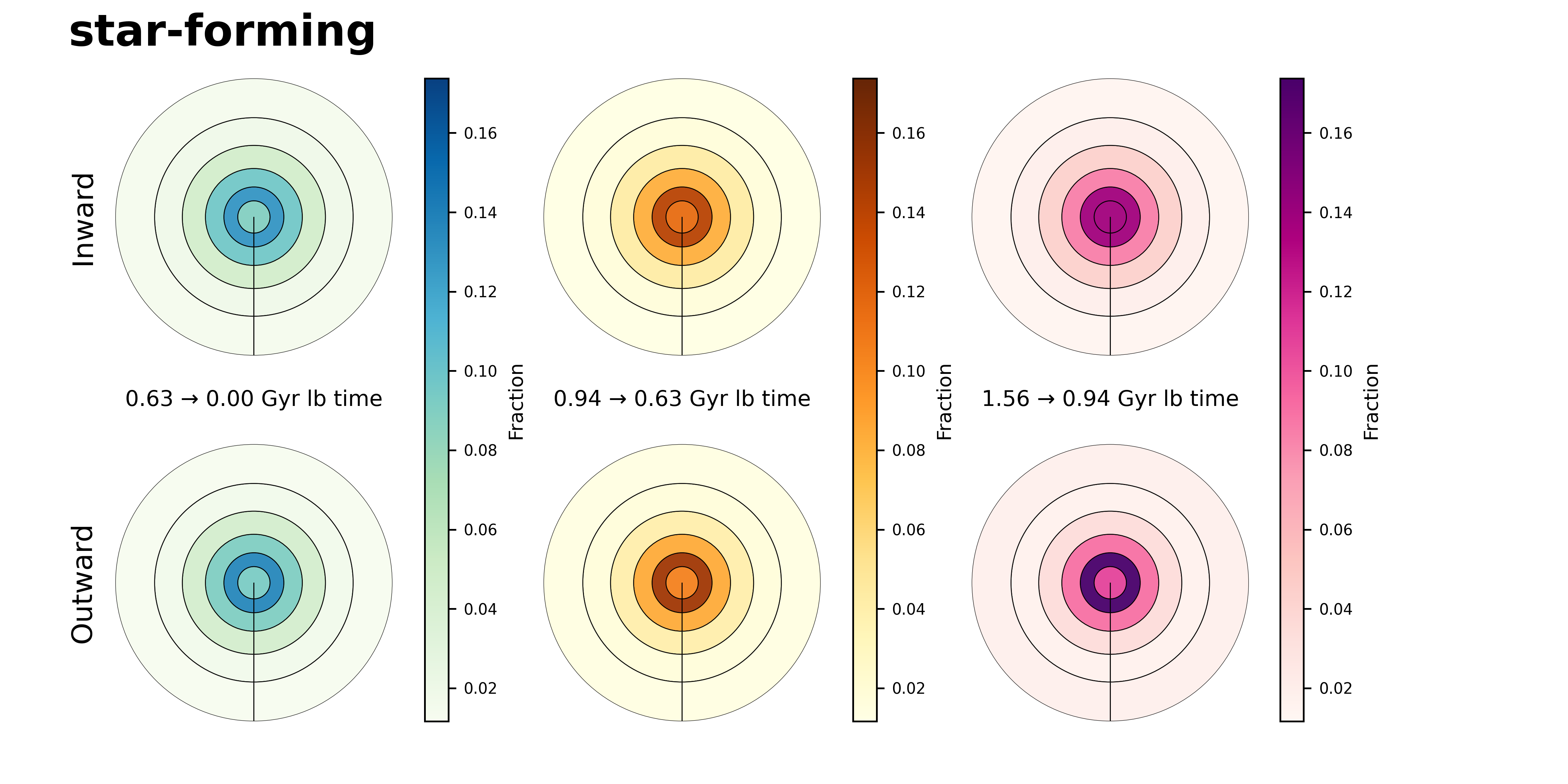}
    \includegraphics[page = 2, width=0.95\linewidth]{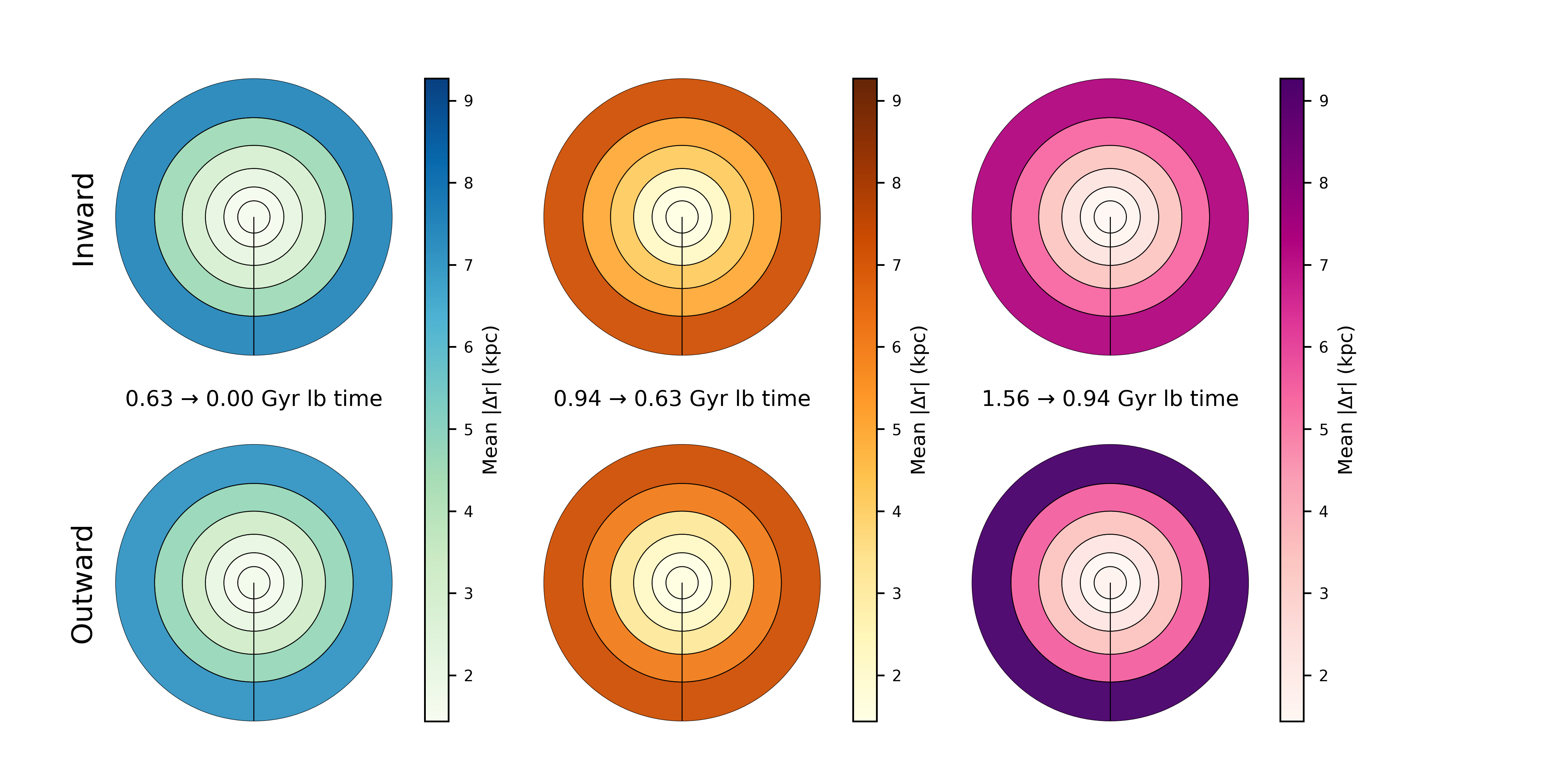}
    \caption{Window 3 (star forming): Similar to Figures~\ref{fig:window1_migration} \& \ref{fig:window2_migration}, there is little movement in and our of the bins with the exception of near the galactic center, and stars that have entered or exited their bins that have traveled the greatest distance lay in the outermost radial bins. For this window the distances traveled are greater than those in the quiescent and decreasing windows ($\leq 5$ kpc vs $\leq 6$ kpc), but these greater distances traveled at greater radii appear to be related to resonances better captured by the fine binning as seen in Figure~\ref{fig:window3_migration_bin_dist_compare}.}
    \label{fig:window3_migration}
\end{figure*}

\noindent The behavior is essentially similar in the star forming phase, presented in Fig.~\ref{fig:window3_migration}, although mean distances traveled by bin-crossing stars increase by $50\%$ in the outermost bin compared to the previous case. This is not enough however to have stars cross more than one bin, and the fraction of stars crossing bins over $\Delta t=0.5$ Gyr remains $<2\%$ in this bin anyway.

\begin{figure*}
    \centering
    \includegraphics[page = 1, width=0.95\linewidth]{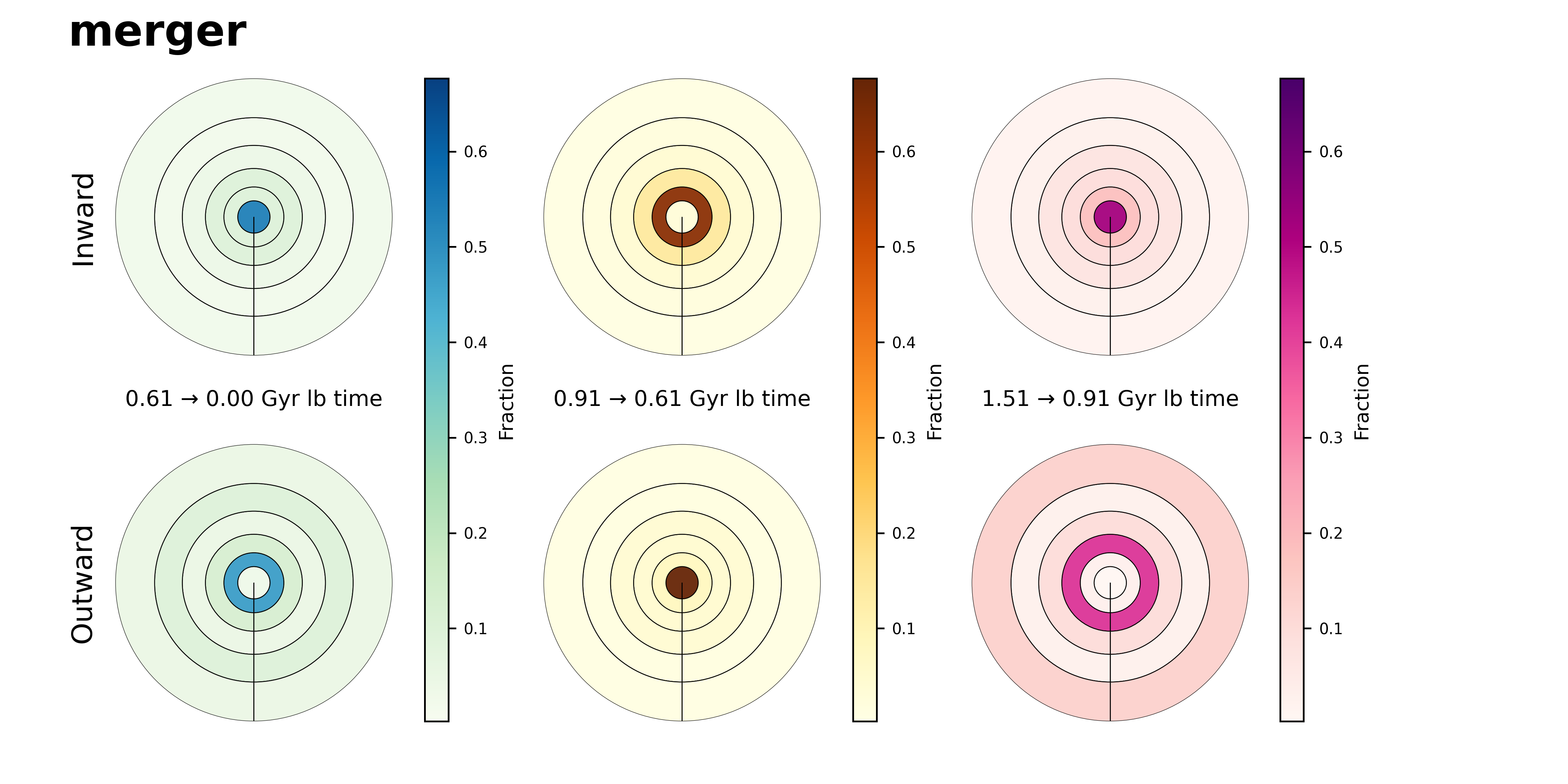}
    \includegraphics[page = 2, width=0.95\linewidth]{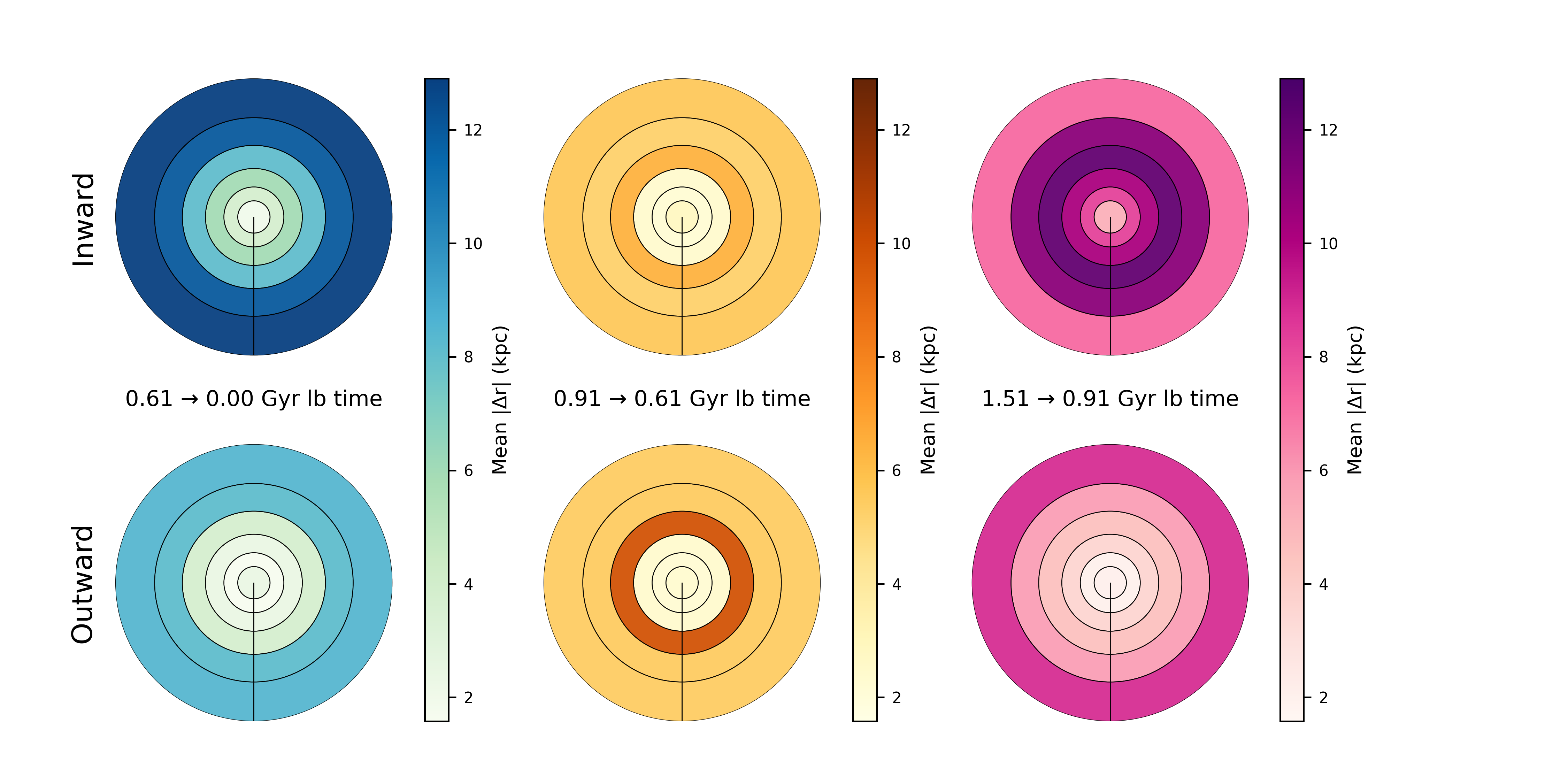}
    \caption{Window 4: The merger window shows that the migration mow spreads over the extent of the galaxy for all bins. For every epoch there are at least 10-20\% of all stars crossing into or out of all radial bins, and the mean distance moved is equal to the radius of the galaxy in some bins. The movement into and out of bins shows the merger in progress, movement of stars spans all radial bins. The fraction of star particles displaced from their original bins now is as much as 70\% and the distances traveled are as far as 13 kpc.}
    \label{fig:window4_migration}
\end{figure*}

\noindent Unsurprisingly, in the merger phase, presented in Fig.~\ref{fig:window4_migration}, deviations become more extreme hence start significantly affecting the quality of the radial profile reconstruction. Mean distances traveled by bin-crossing stars now can reach $13$ kpc inwards including intermediate radial bins. While this is not surprising in the context of structures merging together, this prevents accurate reconstruction of radial profiles of sSFR. As expected, the fractions of bin-crossing stars are also much higher than before, reaching $\sim 60\%$ in central bins and remaining $>10\%$ across radial bins.

\subsection{Across fine regular bins}

We replicate the analysis of Fig.~\ref{fig:window1_migration} in the quiescent phase, this time using regular 1-kpc radial bins. Unsurprisingly, with this fine binning, the fraction of stars migrating across bins within $\Delta t=0.5$ Gyr is much increased, now reaching $20\%$ in the central parts and generally remaining $>5\%$ out to $\sim 6$ kpc from the center of the galaxy. What's more the mean distance traveled by migrating stars almost systematically overshoots 3 times the bin size at $r>10$ kpc and twice the bin size everywhere but in the very center. This enhanced stellar mixing explains the difficulty of fine-binned schemes to maintain accuracy in radial profiles reconstructions of sSFR across large lookback times.

\begin{figure*}
    \centering
    \includegraphics[page = 1, width=0.95\linewidth]{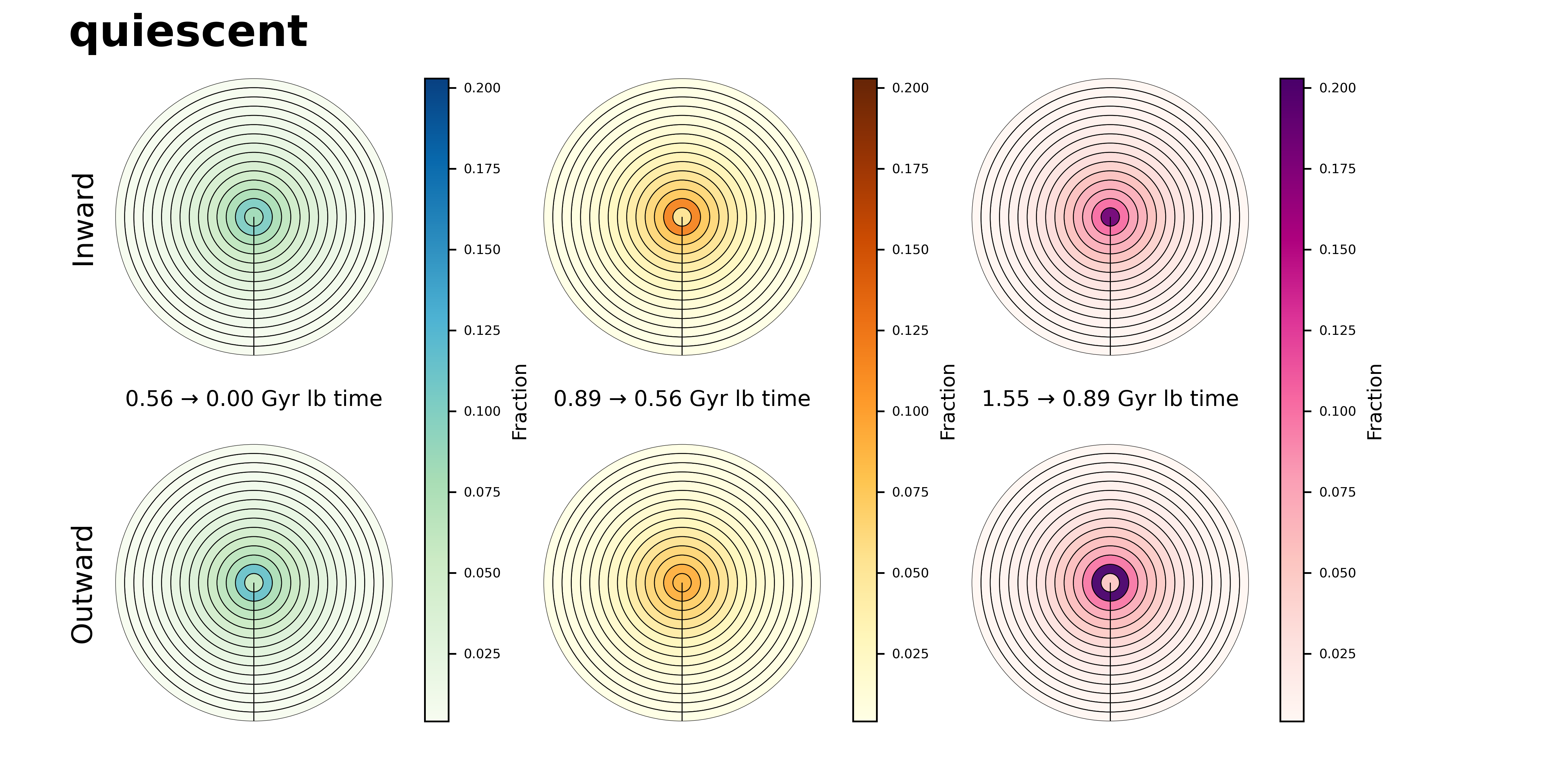}
    \includegraphics[page = 2, width=0.95\linewidth]{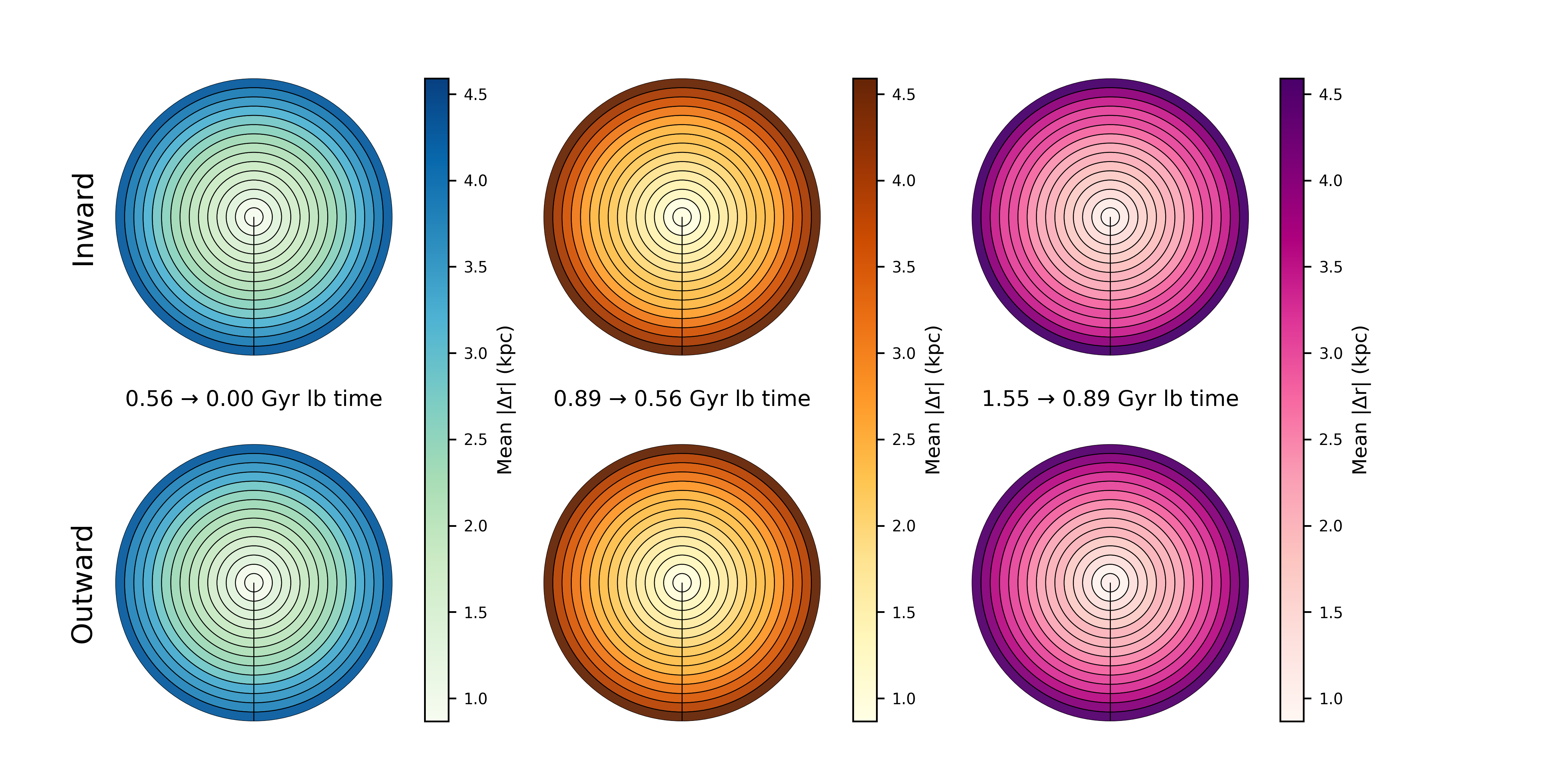}
    \caption{Same as Figure~\ref{fig:window1_migration} but with constant 1 kpc binning. \textbf{Top}: The radial migration between bins by redshift as shown by movement inward or outward for individual stars similarly shows only a small fraction of stars migrating between bins, but stars that have migrated from their original bins and traveled more than 2.5 kpc dominate the last four bins as seen in the \textbf{Bottom} of the figure. The stellar mixing in the central radial bins is consistent with that shown in our observationally motivated variable width bins, and small changes between the bins here should have little impact on the radial profiles.}
    \label{fig:window1_migration_1kpc}
\end{figure*}

\end{document}